\pdfoutput=1 
\documentclass[11pt,fleqn]{article}
\usepackage[utf8]{inputenc}
\usepackage[DIV12]{typearea}
\usepackage{amsmath,amsfonts,amssymb,amscd}
\usepackage{graphicx}
\usepackage{cite}
\usepackage{mathrsfs}
\usepackage{bbm}
\usepackage{units}
\usepackage[small,loose]{subfigure}  
\usepackage{xcolor}
\usepackage{stmaryrd}
\usepackage{pifont}  
\usepackage{hyperref}
\usepackage{cancel}
\usepackage{multicol}
\usepackage{ulem}

\newcommand{\Eqref}[1]{equation~\eqref{#1}}
\newcommand{\Figref}[1]{figure~\ref{#1}}

\newcommand{\Appref}[1]{appendix~\ref{#1}}

\newcommand{\eVdist}{\kern-0.06em}

\newcommand{\GeV}{\text{Ge\eVdist V}}

\newcommand{\TeV}{\text{Te\eVdist V}}
\newcommand{\ev}{\:\text{e\eVdist V}}   

\DeclareMathOperator\arcsinh{arcsinh}

\newcommand{\CenterObject}[1]{\ensuremath{\vcenter{\hbox{#1}}}}

\newcommand{\e}{\mathrm{e}}

\definecolor{darkgreen}{HTML}{109930}

\def\mytitle{Exploring extra dimensions through inflationary tensor modes}
\title{\mytitle}

\begin{document}

\vspace*{1.0cm}

\begin{center}
{\LARGE\textbf{\mytitle}}

\renewcommand*{\thefootnote}{\fnsymbol{footnote}}

\vspace{1.2cm}
\large
\textbf{
Sang Hui Im\footnote[1]{shim@th.physik.uni-bonn.de}{}, 
Hans Peter Nilles\footnote[2]{nilles@th.physik.uni-bonn.de}{},
Andreas Trautner\footnote[3]{atrautner@uni-bonn.de}{}
}\normalsize
\\[5mm]
\textit{Bethe Center for Theoretical Physics \\ and \\ Physikalisches Institut der Universit\"at Bonn,\\
Nussallee 12, 53115 Bonn, Germany
} 
\end{center}
\vspace*{12mm}

\begin{abstract}
Predictions of inflationary schemes can be influenced by the
presence of extra dimensions. This could be of particular
relevance for the spectrum of gravitational waves in models where
the extra dimensions provide a brane-world solution to the
hierarchy problem. Apart from models of large as well as exponentially
warped extra dimensions, we analyze the size of tensor modes in the
Linear Dilaton scheme recently revived in the discussion of
the ``clockwork mechanism". The results are model dependent,
significantly enhanced tensor modes on one side and a
suppression on the other. In some cases we are led to a scheme of
``remote inflation", where the expansion is driven by energies
at a hidden brane. 
In all cases where tensor modes are enhanced, 
the requirement of perturbativity of gravity
leads to a stringent upper limit on the allowed Hubble 
rate during inflation.

\end{abstract}
\thispagestyle{empty}
\clearpage

\section{Introduction}
Combining the presence of extra space dimensions with an inflationary phase
of the early universe might lead to novel and testable insights in cosmology.
Extra dimensions arise in unified schemes like superstring theory and could
provide solutions of the weak scale hierarchy problem. 
Predictions of the inflationary scheme such as e.g.\ the fluctuations in the cosmic microwave
background test the situation at highest cosmological energies and might
be influenced by the presence of extra dimensions. Setting up an inflationary 
scheme in higher dimensions is a challenge as it would have to
provide a solution to the moduli stabilization problem and also explain why
some space dimensions expand while others are fixed in size. 
The question of the nature of the inflaton field (is it a brane or a bulk field) 
would also have to be addressed.

Up to now, work in this direction concentrated on a very specific picture,
which we call the IRB assumption. It assumes that inflation is driven by a field
on our visible brane and assumes that radii of extra dimensions are fixed
by a separate mechanism that does not influence the specific prediction of
the inflationary scheme. Even with this simplified assumption there is an
impact of the presence of the extra dimensions: gravity could have a different strength in
the bulk and influence the size of tensor modes in the inflationary model
under consideration \cite{Giudice:2002vh} (cf.\ also \cite{Langlois:2000ns, Frolov:2002qm}, and \cite[ch.\ 5.1]{Maartens:2010ar} for a review). 
This is particularly interesting in models that try
to solve the weak scale hierarchy by large or warped extra dimensions.
In this case, matter and inflaton field live on our visible infrared (IR) brane while
gravity is stronger in the bulk and at a hidden ultraviolet (UV) brane.\footnote{%
Concerning the terminology in this work, we will always refer to \textit{our} brane - at which the Standard Model lives - as the visible brane and place it
at the origin ($z=0$) of any extra-dimensional coordinate. Irrespectively, we refer to the IR brane as one at which the weak scale hierarchy problem is solved,
and to the UV brane as one at which the hierarchy problem is not solved.
In this sense, the visible brane coincides with the IR brane if our weak scale hierarchy problem is solved by
the presence of an extra dimension.}
Work along these directions has been done in the framework of large extra
dimensions (LED) \cite{ArkaniHamed:1998rs,ArkaniHamed:1998nn}
and warped extra dimensions \`a la Randall-Sundrum (RS) \cite{Randall:1999ee}.

The present work has its origin in the study of inflationary models 
within the so-called Linear Dilaton model (LD) \cite{Antoniadis:2001sw,Antoniadis:2011qw,Cox:2012ee,Baryakhtar:2012wj} 
which regained popularity from a discussion of aligned axions \cite{Kim:2004rp,Choi:2014rja} 
and the clockwork scheme \cite{Choi:2015fiu,Kaplan:2015fuy,Giudice:2016yja,Craig:2017cda,Giudice:2017suc}. 
It can accommodate a solution of the weak-scale hierarchy problem in a braneworld scenario
(with IR- and UV-brane) with power law warping (in contrast to exponential
warping in the RS case). When studying the LD model within the framework
of the simplifying IRB assumption we were led to some inconsistencies to be
explained later. To achieve the standard inflationary picture on the visible 
brane some contributions from the invisible brane (or bulk) are needed. This
observation leads us to reconsider a more general picture of extra-dimensional
inflation beyond the simplest assumption (both in the LD and the RS model) and
this is a main subject of this paper.

Our discussion is organized as follows. In chapter 2 we will summarize the
formulae relevant for the discussion. As a warm-up, we then repeat the
discussion for the LED case with one extra dimension using the simplifying
IRB assumption. In this case we find an enhancement of tensor modes as the
effective Planck mass is reduced through extra dimensional
effects. There is an upper bound on the Hubble scale during inflation as well
as on the reheating temperature after inflation. 
Still in some regions of parameter space one could find a sizable tensor-to-scalar 
ratio due to the transdimensional enhancement. We then turn to the RS model
and consider inflation under the simplifying IRB assumption. 
Again the Planck mass is reduced during inflation, 
implying that the strength of gravity is enhanced.
The tensor-to-scalar ratio is enhanced and we obtain an upper bound on the
Hubble parameter. This reproduces known results in the literature based on
the simplified assumptions \cite{Giudice:2002vh}.

Next we turn our attention to a wider class of inflationary solutions. 
We assume a two-brane RS model with our matter on the IR brane. 
The exponential warping of the extra dimension could explain the weakness of
gravity on the visible brane and thus solve the hierarchy problem. 
Even in the static case we see that the properties of the system strongly depend
on the physics at both branes: IR and UV. The radius of the extra dimension
can be tuned through a choice of the brane tensions on the visible and hidden
brane. The implications of energy on the two branes are highly interdependent.
A model where inflation is driven originally by inflationary dynamics at the
visible brane could be made static by tuning the energy density of the hidden brane.
Alternatively the physics at the hidden brane could be the only source of
inflationary behavior, a phenomenon one might call ``remote'' inflation. 
In this case the Planck mass is enhanced during inflation implying that the
strength of gravity is reduced compared to the static case. 
We also treat a specific case in this general class of solutions 
discussed earlier by Nihei-Kaloper-Kim-Kim (NKKK) \cite{Nihei:1999mt, Kaloper:1999sm, Kim:1999ja} as well as the single 
brane warped model RS2 \cite{Randall:1999vf}.

In chapter \ref{sec:LD} we consider the Linear Dilaton model. 
As in RS we have a negative cosmological constant in the bulk and two branes 
(with matter fields and inflaton field on the visible IR brane). 
A hierarchy of scales appears because of a power-law warping 
(in contrast to exponential warping in the RS-case). 
This situation is more complicated as there is an additional degree of 
freedom in the bulk (the dilaton). 
Despite of this we can derive the solution for the static case ($H=0$) 
in full generality.

Unfortunately this is no longer true in the expanding case $H\neq0$.
There we perform a perturbative expansion in $H^2$. We again adopt the
simplifying IRB assumption that we can ignore specific properties of the
stabilization process (imposed by an external mechanism). The study of
this naive case leads to the amazing result that we are not able to recover
the standard inflationary expansion on the visible brane (in contrast to the
RS-case). The IRB assumption leads to a contradiction. Inflationary
behavior on the visible brane can only be obtained if there is some contribution
from the hidden brane. This is similar to the discussion of ``remote"
inflation in the RS-case. But here we have no choice: 
some ``remote'' contribution is required 
(in contrast to the RS-case where such a contribution was optional). 
The origin of this particular behavior is the presence of the
dilaton as an additional bulk degree of freedom.

On the other hand, the presence of this additional bulk field opens the
possibility to stabilize the radius with the dilaton field without the use
of new external degrees of freedom. This situation is examined in
chapter \ref{sec:LDS}. We again have to do a perturbative expansion in $H^2$
(completed with a full numerical solution). Surprisingly this situation
allows the conventional inflationary scenario where inflation is driven
from the visible brane (with no ``remote'' contribution needed). 
The tensor modes can be computed and are found to be suppressed compared to the
four-dimensional case (in contrast to the RS case). We also compare these
results with the analysis of ref.\ \cite{Kehagias:2016kzt} done in  a conceptually
different setup. Chapter \ref{sec:Conclusions} is devoted to conclusions and outlook.

\section{General considerations}
\subsection{Metric and expansion law}
We are interested in five-dimensional ``braneworld'' scenarios where
gravity is propagating in five dimensions, while the Standard Model
is confined to a four-dimensional slice of spacetime. The weak scale hierarchy problem can then be solved by
an apparent, large four-dimensional Planck mass which is caused by the tiny overlap 
of the (massless) graviton zero mode with the visible brane.

In a very general manner, the action is given by
\begin{equation}\label{eq:action}
 S~=~\int\mathrm{d}^4x\,\mathrm{d}z\,\sqrt{|g|}\left\{-\frac{M^3}{2}\mathcal{R} + \mathcal{L}_b(x,z) + \frac{1}{\sqrt{|g_{55}|}}\left[\mathcal{L}_0\,\delta(z)+\mathcal{L}_\pi\,\delta(z-\pi\,R)\right] \right\}\;.
\end{equation}
For a realistic cosmology, including a flat four-dimensional space, the most general ansatz for the metric can be written as
\begin{equation}\label{eq:generalMetric}
 \mathrm{d}s^2~=~n(z,t)^2\,\mathrm{d}t^2-A(z,t)^2\,\delta_{ij}\,\mathrm{d}x^i\mathrm{d}x^j-b(z,t)^2\,\mathrm{d}z^2\;.
\end{equation}
In hindsight of the properties of our anticipated solutions, however, we choose a simpler form of the metric as ansatz:
\begin{equation}\label{eq:ansatzMetric}
 \mathrm{d}s^2~=~f(z)^2\,\left(\mathrm{d}t^2-a(t)^2\,\delta_{ij}\,\mathrm{d}x^i\mathrm{d}x^j\right)-\mathrm{d}z^2\;.
\end{equation}
The assumptions which allow for a reduction of \eqref{eq:generalMetric} to the simpler form \eqref{eq:ansatzMetric}
are
\begin{itemize}
 \item[(i)] $\dot b(z,t)=0$ (the size of the extra dimension is static), and 
 \item[(ii)] $A(z,t)=f(z)\,a(t)$ ($A$ is a separable function).
\end{itemize}
The first assumption (i) is certainly fulfilled if there is a mechanism to stabilize the size of the extra dimension, 
for example via a stabilizing potential for the radion mode \cite{Goldberger:1999uk}.
Alternatively, if cosmological constants are the only form of energy density on the branes, 
one can achieve a consistent solution for a static extra dimension also by fine-tuning of the brane energy densities (c.f.\ e.g.\ \cite{Nihei:1999mt, Kaloper:1999sm, Kim:1999ja}).
In either case, a stabilization will involve a contribution to the $(55)$ component of the energy-momentum tensor $T_{MN}$.\footnote{%
Our conventions are: $M,N,..=0,1,2,3,5$; $\mu,\nu,..=0,1,2,3$; $i,j,..=1,2,3$; metric signature $(+1,-1,-1,-1,-1)$; dots and primes denote the derivatives with respect to $t$ or $z$, respectively.}
The $(55)$ component of the Einstein equations, thus, serves to determine the size of the 5th dimension
independently of the details of the stabilization mechanism \cite{Csaki:1999mp, Kanti:1999sz, Kanti:2000rd}. 
This means that, as long as the stabilization mechanism decouples from all other equations, 
we can simply put aside the $(55)$ equation while assuming that the radius is stabilized at some value (see e.g.\ \cite{Giudice:2002vh}). 
We will later see that this is not always the case, 
and we will then also take into account the $(55)$ equation.
The remaining choice $b(z)=1$ then corresponds to a choice of coordinate system.

One can show that under the assumption (i), point (ii) is fulfilled if and only if 
$(n/A)$ is independent of $z$. Considering matching conditions on the four-dimensional branes, this requires that $\mathcal{L}_{0,\pi}$ is time independent \cite{Lesgourgues:2000tj}, i.e.\ the energy densities are dominated by cosmological constants. 
This is a good assumption here, because we are interested in inflationary solutions of the scale factor $a(t)$.

The Einstein's equations are of the form
\begin{equation}\label{eq:Einstein}
 G_{MN}~=~\mathcal{R}_{MN}-\frac{\mathcal{R}}{2}\,g_{MN}~=~\kappa^2\,T_{MN}\;,
\end{equation}
with $\kappa^2\equiv M^{-3}$. With the ansatz \eqref{eq:ansatzMetric} the following features arise:
\begin{itemize}
 \item The $(00)$ and $(ij)$ equations are degenerate.
 \item The $(05)$ equation is automatically fulfilled. Note that in the more general ansatz \eqref{eq:generalMetric},
 the $(05)$ equation gives rise to the insight that
 \begin{equation}
  \frac{\dot{A}(z,t)}{n(z,t)}~=~\alpha(t)\;,
 \end{equation}
 is independent of $z$ \cite{Kanti:1999sz}. Due to the assumption (ii) of our ansatz above, however, this relation is automatically fulfilled here.
\end{itemize}
The only relevant Einstein equation, hence, is the $(00)$ component of \eqref{eq:Einstein} which is given by
\begin{equation}\label{eq:GeneralMetricWarpingLEDRS}
 -3\,f^2\left\{ \frac{f''}{f}+\frac{f'^2}{f^2}\right\}+3\,\frac{\dot{a}^2}{a^2}~=~\kappa^2\,T_{00}\;.
\end{equation}
Assuming the bulk and brane Lagrangians to be time independent it readily follows from \eqref{eq:GeneralMetricWarpingLEDRS} that 
\begin{equation}
H~:=~\frac{\dot{a}}{a}\;,
\end{equation}
is constant. Upon integration 
\begin{equation}
a(t)~=~\e^{H\,t}\;,
\end{equation}
and we realize that $H$ corresponds to the physical expansion rate of a three-dimensional slice of space 
\textit{at the five dimensional point} $z_0$ \textit{where} $f(z_0)$=1. The proper physical Hubble rate at a different 
slice of four-dimensional space time, say at $z=z_1$, can be obtained from $H$ by a redefinition 
of the time coordinate $f^2(z_1)\mathrm{d}t^2\to\mathrm{d}\tau^2$, and, therefore, is given by
\begin{equation}
 H_{z_1}~=~\frac{H}{f(z_1)}\;.
\end{equation}

\subsection{Effective Planck mass during inflation}
The relevant quantity for the actual strength of 4D gravity is the prefactor of the four-dimensional Ricci scalar (the normalization factor of the zero-mode graviton) which arises upon integrating out
the 5th dimension in \eqref{eq:action}. 
The effective 4D Planck mass $M_{\mathrm{Pl},\mathrm{eff}}$ obtained in this way is given by
\begin{equation}\label{eq:EffectivePlanckMass}
 M_{\mathrm{Pl},\mathrm{eff}}^2~=~M^3\,\int_{\mathrm{5D}}\,\mathrm{d}z\,f(z)^2\;.
\end{equation}
Alternatively, one can deduce a four-dimensional Planck mass $M_{\mathrm{Pl,exp}}$ from the expansion law 
experienced on the visible brane
\begin{equation}\label{eq:expansionLaw}
H^2~=~\frac{\rho}{3\,M_{\mathrm{Pl,exp}}^2}\;. 
\end{equation} 
Here $\rho$ is the approximately constant energy density that drives inflation.
In the picture of the IRB assumption, this energy density is due to a (sufficiently flat) inflaton potential on the visible brane.

The crucial point is that $M_{\mathrm{Pl},\mathrm{eff}}$ and $M_{\mathrm{Pl,exp}}$ will, in general, not 
coincide during inflation, thus, giving rise to a change of the tensor mode power spectrum as compared
to inflation in the purely four dimensional case \cite{Giudice:2002vh}.

\subsection{Tensor modes in braneworld inflation}
As gravitational tensor modes are intrinsically bulk degrees of freedom, they are 
susceptible to the five dimensional geometry during inflation.
For the treatment of tensor mode perturbations (primordial gravitational waves) we follow \cite{Giudice:2002vh} 
(see also \cite{Langlois:2000ns,Frolov:2002qm}). 
Even though the discussion about the tensor modes in \cite{Giudice:2002vh} is based on the IRB assumption that the inflaton field is confined to the visible brane, 
we remark that the result is applicable to general braneworld inflation including ``remote inflation" where inflation is driven by an inflaton field located at the hidden brane. 
This is because the form of the zero mode graviton solution is independent of the specific dynamics responsible for inflation. 
How the spectrum of gravitational waves is modified then depends only on the underlying geometry
and not on the microscopic details of the four-dimensional model of inflation.

The power spectrum of primordial tensor modes 
\begin{equation} \label{Tmode}
 \mathcal{P}_\mathrm{T}(\ell)~=~\frac{2}{\pi^2}\,\left(\frac{H(\ell)}{M_\mathrm{Pl,eff}}\right)^2\;,
\end{equation}
generally deviates from its four-dimensional value. This can be attributed to a 
change of the effective reduced Planck mass $M_\mathrm{Pl,eff}$ during inflation \cite{Giudice:2002vh}.
Kaluza-Klein modes other than the massless tensor mode are not relevant
because they are separated by a sufficiently broad mass gap 
\cite{Frolov:2002qm}. 

In sharp contrast, scalar mode (density) perturbations originating from quantum fluctuations of the inflaton field 
do depend on the specific scenario of braneworld inflation. 
If one adopts the IRB assumption, the scalar perturbation is a purely four-dimensional degree of freedom confined to the visible brane. Then,  
given the ordinary four-dimensional Hubble law \eqref{eq:expansionLaw} on the brane, 
the presence of extra dimensions does (to leading order in slow-roll) not affect the power spectrum of scalar metric perturbations which is given by \cite{Giudice:2002vh}
\begin{equation}
 \mathcal{P}_\mathrm{S}(\ell)~=~\frac{1}{8\,\pi^2}\,\frac{1}{\epsilon}\,\left(\frac{H(\ell)}{M_\mathrm{Pl,exp}}\right)^2\;.
\end{equation}
Consequently, the tensor-to-scalar ratio in scenarios with an extra dimension is modified to
\begin{equation}
 \frac{\mathcal{P}_\mathrm{T}(\ell)}{\mathcal{P}_\mathrm{S}(\ell)}~\approx~\frac{M_\mathrm{Pl,exp}^2}{M_\mathrm{Pl,eff}^2}\times\left.\frac{\mathcal{P}_\mathrm{T}(\ell)}{\mathcal{P}_\mathrm{S}(\ell)}\right|_\mathrm{4D}\; ,
\end{equation}
if the IRB assumption is adopted. For other cases including ``remote inflation", 
we could not make a definite statement on the tensor-to-scalar ratio at the moment. 
However, we still predict the altered behavior of primordial tensor modes based on the knowledge of the effective Planck mass during inflation.

\section{Large Extra Dimensions and Randall-Sundrum scenario}
\subsection{General form of the metric warping}
The actions of the LED model \cite{ArkaniHamed:1998rs,ArkaniHamed:1998nn} and of the Randall-Sundrum model \cite{Randall:1999ee} 
are given by simplifications of \eqref{eq:action}.
Let us consider the case of a bulk cosmological constant in addition to the unspecified radion stabilization mechanism in the bulk 
$\mathcal{L}_b(x,z)~=~-\Lambda+\mathcal{L}_{\mathrm{Rad}}(z)$. 
Then, the $(00)$ Einstein equation \eqref{eq:GeneralMetricWarpingLEDRS} which determines the metric warping $f$ is given by
\begin{equation}\label{eq:00inBulk}
\left(f^2\right)''-2\,\frac{\dot{a}^2}{a^2}+\frac{2}{3}\,\kappa^2\,\Lambda\,f^2~=~0\;,
\end{equation}
locally \textit{in the bulk}. Using the definition of $H$ as well as 
\begin{equation}
\mu^2~:=~-\frac{2}{3}\,\kappa^2\,\Lambda\;,
\end{equation}
the $(00)$ equation in the bulk reads
\begin{equation}\label{eq:General00inBulk}
\left(f^2\right)''-2\,H^2-\mu^2\,f^2~=~0\;.
\end{equation}
Depending on the sign of $\Lambda$, this equation has the general solutions (cf.\ also \cite{Binetruy:1999hy})
\renewcommand{\arraystretch}{1.45}
\begin{align}\label{eq:00GenSol}
 f^2(z)~=~\left\{\begin{array}{lll} H^2\,z^2+c_1\,z+c_2 & & \Lambda~=~0\;, \\
 A\,\mathrm{e}^{-\mu\,z}+B\,\mathrm{e}^{\mu\,z}-\frac{2\,H^2}{\mu^2} & \qquad\text{for}\qquad & \Lambda~<~0\;, \\
 C\,\sin{\mu\,z}+D\,\cos{\mu\,z}+\frac{2\,H^2}{\mu^2} & & \Lambda~>~0\;, \\ \end{array}\right.
\end{align}
where we take $\mu>0$ without loss of generality.
Each solution has two constants which are to be determined from the boundary conditions.
Depending on the setting, the boundary conditions are set by symmetry constraints or the placement of branes.
For example, if there is an infinitely thin brane with constant energy density (``brane tension'')
at a position $z=z_0$, the four-dimensional ``brane'' Lagrangian takes the form
\begin{equation}
 \mathcal{L}_{z_0}(x)\,\delta(z_0)~=~-\rho_{z_0}\,\delta(z_0)\;.
\end{equation}
This gives rise to a discontinuity of the first derivative of $f$ across the brane \cite{Israel:1966,Binetruy:1999ut} which is given by 
\begin{equation}\label{eq:JumpCondition}
 \frac{f'(z_0^+)-f'(z_0^-)}{f(z_0)}~=~-\frac{\kappa^2}{3}\,\rho_{z_0}\;.
\end{equation}
Solutions for $f$ in regions which are separated by branes are a priori unrelated.
Relations between solutions for $f$ in the different regions can be obtained, for example, by requiring that the functions are related by the 
orbifold transformation $z\to-z$, or similar relations.

\subsection{Solution for the metric during inflation}
\subsubsection{Large Extra Dimensions}
The LED case is characterized by $\Lambda=0$ and the introduction of a single brane.
The extra dimension is taken to be compact, with a size denoted by $2\pi R$.
We follow \cite{Giudice:2002vh} and place the visible brane with tension $\rho_0=\rho$ at $z=z_0=0$.
The boundary condition $f(0)=f(2\pi R)$ fixes $c_1$, 
while the jump condition for the derivatives \eqref{eq:JumpCondition} evaluated at $z_0^+=0$ and 
$z_0^{-}=2\pi R$ fixes the relation
\begin{equation}
 c_2~=~H^2\,\frac{6\,\pi\,R}{\kappa^2\,\rho}\;.
\end{equation}
Recall that $H$ corresponds to the usual four-dimensional Hubble rate at the slice where $f(z)=1$.
Since we are interested in the visible brane at $z=0$ we chose the four-dimensional coordinates such as to normalize $f(0)=1$
corresponding to $c_2=1$. 

Altogether, the solution then is given by
\begin{equation}\label{eq:LEDSolution}
 f^2(z)~=~H^2\left(z^2-2\,\pi\,R\,z\right)+1\;,\qquad\text{with}\qquad H^2~=~\frac{\kappa^2\,\rho}{6\,\pi\,R}\;.
\end{equation}

Finally we can obtain the effective Planck mass in 4D from the metric warping and compare it to the effective Planck mass appearing in 
the expansion law. From the metric warping one finds 
\begin{equation}\label{eq:EffectivePlanckMassLED}
  M_{\mathrm{Pl},\mathrm{eff}}^2~=~M^3\,\int_0^{2\,\pi\,R}\,\mathrm{d}z\,f(z)^2~=~M^3\,2\,\pi\,R\,\left(1-\frac23\,\pi^2\,R^2\,H^2\right)\;.
 \end{equation}
This shows that there is an $H$ dependent correction to the perceived strength of gravity in four dimensions as compared to the static case. 

In contrast, from the four-dimensional Hubble expansion law one finds
 \begin{equation}
  H^2~=~\frac{\kappa^2\,\rho}{6\,\pi\,R}
	~\equiv~\frac{\rho}{3\,M_{\mathrm{Pl,exp}}^2}\;\quad\Rightarrow\quad M_{\mathrm{Pl,exp}}^2~=~M^3\,2\,\pi\,R~=~\left. M_{\mathrm{Pl,eff}}^2\right|^{\mathrm{LED}}_{H=0}\;.
 \end{equation}
We see that the differently derived four-dimensional Planck masses deviate during inflation.
While $M_{\mathrm{Pl,exp}}$ is the relevant scale for perturbations of the scalar mode, $M_{\mathrm{Pl},\mathrm{eff}}$ is the relevant scale for the tensor mode perturbations.
Consequently the tensor-to-scalar ratio is modified due to transdimensional effects \cite{Giudice:2002vh}.

The fact that the effective Planck mass in \eqref{eq:EffectivePlanckMassLED} is reduced shows that gravity is stronger during inflation, 
i.e.\ the tensor-to-scalar ratio is enhanced. In fact, if the Hubble scale is too large the metric warp function \eqref{eq:LEDSolution}
crosses zero at which point there appear curvature singularities in the bulk.\footnote{%
We thank the referee for drawing our attention to this point.}
This signals the breakdown of perturbative gravity. Avoiding this situation results in an
upper bound 
\begin{equation}\label{eq:LEDBound}
 \pi^2\,R^2\,H^2~<~1\;.
\end{equation}
This result generalizes to $n$ compact extra dimensions in the form $c_n\pi\,R\,H<1$, with a factor $c_n\sim\mathcal{O}(1)$ that depends on the details of the compactification \cite{Giudice:2002vh}.
Taking all 5D scales to be $M\sim\mu\sim\Lambda\sim\TeV$ and requiring a successful solution to the hierarchy problem one finds
$R\sim10^{30/n-19}\,\mathrm{m}$ resulting in an approximate upper bound on $H<10^{-30/n}\,\TeV$ corresponding to a maximal reheating temperature
of $T_\mathrm{RH}<10^{21/2-15/n}\,\GeV$. For the specific case of $n=2$ this corresponds to $H<10^{-12}\,\GeV$ and $T_\mathrm{RH}<10^{3}\,\GeV$.

Depending on the assumed microscopic model of inflation there will eventually be even stronger bounds imposed on the product $HR$ by the (non-)observation
of CMB $B$-mode polarization.

\subsubsection{Randall-Sundrum scenarios}

The Randall-Sundrum scenario \cite{Randall:1999ee} is characterized by taking $\Lambda<0$ while introducing two branes.
Without loss of generality, the branes are placed at $z=0$ and $z=\pi R$. The respective brane tensions are denoted by $\rho_0$ and $\rho_\pi$, respectively. 
We will solve for $f(z)$ in the region $0\leq z\leq \pi R$,
while the solution in the region $-\pi R\leq z\leq 0$ can be obtained by the orbifold transformation $z\to-z$.

The general solution for $f(z)^2$ in the bulk is given by (cf.\ \eqref{eq:00GenSol})
\begin{equation}\
f^2(z)~=~A\,\mathrm{e}^{-\mu\,z}+B\,\mathrm{e}^{\mu\,z}-\frac{2\,H^2}{\mu^2}\;,
\end{equation}
with two constants $A$ and $B$ that will shortly be fixed by boundary conditions.
The effective four-dimensional Planck mass can be computed in a general fashion,
\begin{equation}
 M_{\mathrm{Pl},\mathrm{eff}}^2~=~2\,M^3\,\int_0^{\pi\,R}\,\mathrm{d}z\,f(z)^2~=~\frac{2\,M^3}{\omega\,\mu}\left(1-\omega\right)\left[B+\omega\,A-\frac{\omega}{(1-\omega)}\frac{2\,H^2\,\pi\,R}{\mu}\right]\;,
\end{equation}
where we have introduced the warp factor of the static case
\begin{equation}
\omega:=\mathrm{e}^{-\mu\,\pi R}. 
\end{equation}

Let us compute $A$, $B$, and $H$ for general (time-independent) boundary conditions.
The junction conditions \eqref{eq:JumpCondition} take the form
\begin{equation}\label{eq:RSJumpConditions}
 \frac{f'(0)}{f(0)}~=~-\frac{\kappa^2}{6}\,\rho_0\;,\qquad\text{and}\qquad\frac{f'(\pi R)}{f(\pi R)}~=~\frac{\kappa^2}{6}\,\rho_\pi\;.
\end{equation}
It is useful to define the dimensionless quantities
\begin{equation}
 \lambda_0~:=~\frac{\kappa^2\,\rho_0}{3\,\mu}~=~\frac{\rho_0}{\sqrt{-6\,M^3\,\Lambda}}\;,\qquad\text{and}\qquad\lambda_\pi~:=~\frac{\kappa^2\,\rho_\pi}{3\,\mu}~=~\frac{\rho_\pi}{\sqrt{-6\,M^3\,\Lambda}}\;.
\end{equation}
Obviously, $\lambda_0$ and $\lambda_\pi$ compare the brane tensions to the bulk cosmological constant. In particular, we will shortly see that for $\lambda_0=-\lambda_\pi=\pm1$ the originally considered static case \cite{Randall:1999ee} is obtained. 
Altogether the constraints from the boundary conditions can be written as 
\begin{align}\nonumber\label{eq:BCRS}
 f_0^2~&=~A+B-\frac{2\,H^2}{\mu^2}\;,& f_\pi^2~&=~A\,\omega+B\,\omega^{-1}-\frac{2\,H^2}{\mu^2}\;,& \\ 
 \lambda_0\,f_0^2~&=~A-B\;,& -\lambda_\pi\,f_\pi^2~&=~A\,\omega-B\,\omega^{-1}\;,& 
\end{align}
where we use the abbreviations 
\begin{equation}
f_0:=f(0)\;,\quad f_\pi:=f(\pi R)\;.
\end{equation}
The first line of \eqref{eq:BCRS} are simple identities, while the second line arises from the 
junction conditions \eqref{eq:RSJumpConditions}. 
Taking $\mu$, $\lambda_0$, $\lambda_\pi$, $R$ together with a normalization condition for $f$ (e.g.\ $f_0=1$) as input, 
this can be viewed as a system of four equations with four unknowns
($A$, $B$, $H$, $f_\pi$).
In particular, the expansion rate $H$ is a function of the input parameters and completely determined by the requirement
that the jumps in the derivative of $f(z)$ at $z=0$ and $z=\pi R$ are consistent with the imposed brane tensions.

Let us take the visible brane to be located at $z=0$ and, therefore, adopt the normalization
$f_0=1$. A general expression for the Hubble rate (at $z=0$) then is given by
\begin{equation}\label{eq:ExpansionRate}
 H^2~=~\frac{\mu^2}{2}\,\frac{(1-\lambda_0)(1-\lambda_\pi)-\omega^2\,(1+\lambda_0)(1+\lambda_\pi)}{\lambda_\pi(1-\omega)^2+\omega^2-1}\;,
\end{equation}
while the parameters $A$ and $B$ can be expressed as
\begin{equation}\label{eq:Coefficients}
 A~=~\frac12(1+\lambda_0)+\frac{H^2}{\mu^2}\;,\qquad\text{and}\qquad B~=~\frac12(1-\lambda_0)+\frac{H^2}{\mu^2}\;.
\end{equation}
Let us reproduce some known results, while pointing out novel insights.

\paragraph{Static RS1.}
For an arbitrary but fixed radius $R$ (i.e.\ a fixed warp factor $\omega$) we see that the expansion rate \eqref{eq:ExpansionRate}
vanishes if $\lambda_0=-\lambda_\pi=\pm1$. This corresponds to the cases $(A=1, B=H=0)$ as well as $(A=H=0, B=1)$ implying that the metric 
takes the standard form
\begin{equation}\label{eq:RSMetric}
 \mathrm{d}s^2~=~\mathrm{e}^{\mp\mu z}\,\left(\mathrm{d}t^2-\delta_{ij}\,\mathrm{d}x^i\mathrm{d}x^j\right)-\mathrm{d}z^2\;.
\end{equation}
Clearly, this is the originally considered RS1 case \cite{Randall:1999ee}, which is static.
For the case $\lambda_0=-\lambda_\pi=-1$ we reproduce the well-known result 
\begin{equation}\label{eq:MPlstaticRS1}
 \left.M^2_\mathrm{Pl,eff}\right|^{\mathrm{RS1}}_{H=0}~=~\frac{2\,M^3}{\omega\,\mu}\left(1-\omega\right)~=~\frac{2\,M^3}{\mu}\left(\mathrm{e}^{\mu\,\pi R}-1\right)\;,
\end{equation}
finding that $M_{\mathrm{Pl},\mathrm{eff}}$ appears exponentially enhanced over the fundamental scale $M$.

Note that tuning $\lambda_0$ and $\lambda_\pi$ to equal but opposite values $\pm1$ is \textit{not} the only possibility in order to obtain a static ($H=0$) case.
Alternatively, there can be a non-trivial interplay of the size of the fifth dimension
and the expansion rate of the four-dimensional slices.
\begin{figure}
\centering
 \includegraphics[width=0.5\textwidth]{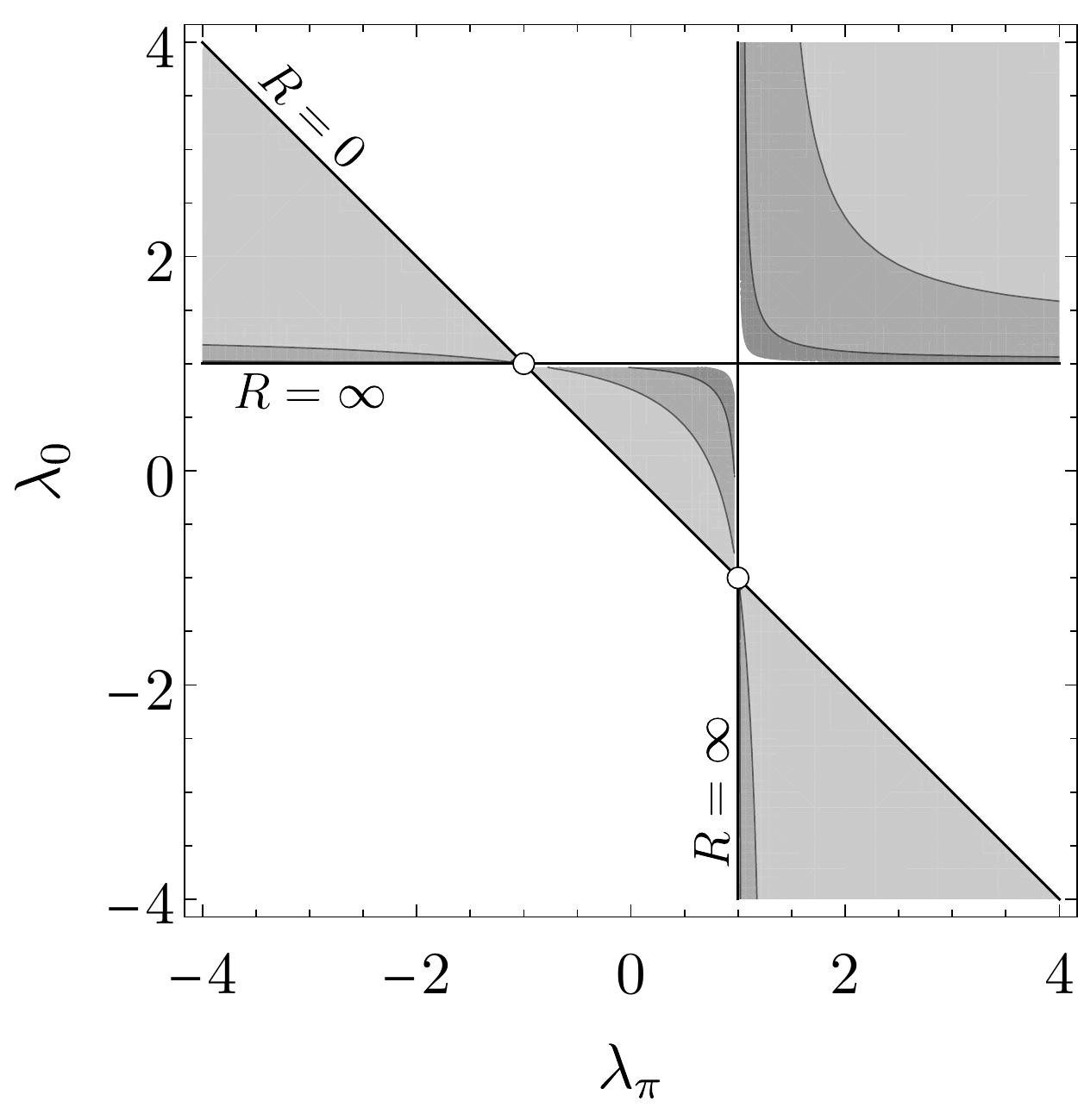}
\caption{Allowed regions (filled) for the normalized brane tensions $\lambda_0,\lambda_\pi$. 
The contours illustrate the value of $R=R_\mathrm{RS}$ (darker means bigger) such that $H=0$. 
At the solid boundary lines $R$ asymptotes the values indicated in the plot.
Values $\mu\pi R_\mathrm{RS}\approx70$, for which the desired hierarchy between $M_{\mathrm{Pl},\mathrm{eff}}$ and $M$ is obtained, lie very close to 
the vertical and horizontal boundary lines.
Empty circles show the choice of brane tensions in the original Randall-Sundrum scenario, where $H=0$ while $R$ at these points can take any value.}
\label{Fig:RS}
\end{figure}
For a large number of combinations of $\lambda_0$ and $\lambda_\pi$, see \Figref{Fig:RS}, it is possible to tune $R$ to the 
very specific value 
\begin{equation}\label{eq:SaticRadius}
 R_\mathrm{RS}~=~\frac{1}{2\,\mu\,\pi}\,\ln\left[\frac{(\lambda_0+1)(\lambda_\pi+1)}{(\lambda_0-1)(\lambda_\pi-1)}\right]\;,
\end{equation}
which gives rise to a vanishing expansion rate of all four-dimensional slices. 
Most notably, for any given non-trivial value of $\lambda_0\neq\pm1$ and any value of the radius one can always 
tune the tension of the other brane such as to stop inflation.

This clearly demonstrates that for inflationary solutions the evolution of the IR and UV branes are highly interdependent.
In the most extreme case, for example, one could  drive inflation of the visible brane by dynamics located solely on the hidden 
brane (``remote'' inflation).

\paragraph{Expanding RS1.}
Using the general results \eqref{eq:ExpansionRate} and \eqref{eq:Coefficients}, we can also reproduce the inflationary case considered by Giudice et al.\ \cite{Giudice:2002vh}. Therefore,
we take $\lambda_0=-1+\epsilon_0$ and $\lambda_\pi=1$. This corresponds to the usual fine tuned brane tensions 
of the static case plus an extra energy density $\rho_0$ on the visible brane, characterized by the dimensionless parameter
\begin{equation}\label{eq:epsilonDef}
 \epsilon_0~:=~\frac{\kappa^2\,\rho_0}{3\,\mu}\;.
\end{equation}
The Hubble rate comes out as
\begin{equation}\label{eq:HRS}
 H^2~=~\frac{\epsilon_0\,\mu^2}{2}\,\frac{\omega}{1-\omega}~=~\frac{\kappa^2\,\mu\,\rho_0}{6}\,\frac{\omega}{1-\omega}~=~\frac{\rho_0}{3\,\displaystyle{\left.M^2_\mathrm{Pl,eff}\right|^{\mathrm{RS1}}_{H=0}}}\;,
\end{equation}
and the metric warping is given by 
\begin{equation}\label{eq:WarpFactorRS}
f^2(z)~=~\frac{H^2}{\mu^2\,\omega}\,\mathrm{e}^{-\mu\,z}+\left[1+\frac{H^2}{\mu^2}\left(\frac{2\,\omega-1}{\omega}\right)\right]\,\mathrm{e}^{\mu\,z}-\frac{2\,H^2}{\mu^2}\;. 
\end{equation}
The effective Planck mass during inflation then is
\begin{equation}\label{eq:MplEffRS}
M_{\mathrm{Pl},\mathrm{eff}}^2~=~\left.M_{\mathrm{Pl},\mathrm{eff}}^2\right|^{\mathrm{RS1}}_{H=0}\times\left[1-\frac{H^2}{\mu^2}\left(\frac{1}{\omega}-3+\frac{\omega}{1-\omega}\,2\,\pi\,R\,\mu\right)\right]\;.
\end{equation}
This result is exact. As shown in \Appref{App:RS}, our results for the expansion rate, metric warping and Planck mass agree with the results of \cite{Giudice:2002vh}
after taking into account the different conventions.

Clearly, the Planck mass is reduced during inflation, implying that the strength of gravity is enhanced.
If the Hubble rate becomes too large, $f^2(z)$ crosses zero and there appears a curvature singularity in the bulk. 
Avoiding the onset of strongly coupled gravity thus imposes an upper bound on the Hubble rate
\begin{equation}
H^2~<~\mu^2\,\frac{\omega}{\left(1-\omega\right)^2}~\approx~\mu^2\,\omega\;,
\end{equation}
which also ensures that $M_{\mathrm{Pl},\mathrm{eff}}^2>0$.
Taking all 5D scales to be $M\sim\mu\sim\Lambda\sim\TeV$ and requiring a solution to the hierarchy problem due to a warping $\mu\pi R\sim 70$
this bound restricts the Hubble rate to $H<10^{-12}\,\GeV$. Assuming maximally efficient reheating, this corresponds to a bound $T_\mathrm{RH}<10^{3}\,\GeV$.
The fact that gravity is stronger during inflation generically leads to an enhancement of tensor mode perturbations.

As an interesting alternative, note that one could, in principle, also drive inflation from a completely remote sector that gives 
rise to an approximately constant additional energy density $\rho_\pi$ on the hidden brane. To model this we take $\lambda_0=-1$, $\lambda_\pi=1+\epsilon_\pi$, where 
\begin{equation}\label{eq:epsilonPiDef}
 \epsilon_\pi~:=~\frac{\kappa^2\,\rho_\pi}{3\,\mu}\;.
\end{equation}
The Hubble rate of the visible brane comes out as
\begin{equation}
H^2~=~\frac{\mu^2\,\epsilon_\pi}{2\,\omega\,(1-\omega)-\epsilon_\pi\,(1-\omega)^2}\;,
\end{equation}
and we have to restrict $\epsilon_\pi<2\omega\ll1$ to ensure that $H^2>0$. 
In the case $\epsilon_\pi\ll\omega$ one finds.
\begin{equation}
 H^2~\approx~\frac{\mu^2\,\epsilon_\pi}{2\,\omega}~\approx~\frac{1}{\omega^2}\,\frac{\rho_\pi}{3\,\displaystyle{\left.M^2_\mathrm{Pl,eff}\right|^{\mathrm{RS1}}_{H=0}}}\;.
\end{equation}
This shows that the physical Hubble rate of the visible brane is highly susceptible to even smallest energy densities on the hidden brane. 
For example, the currently observed Hubble rate of $10^{-32}\ev$ on the visible brane can be caused by an additional energy density of 
only $10^{-70}\ev^4$ on the hidden brane. 
The necessary fine-tuning of energy density on the hidden brane demonstrates that the cosmological constant problem of our visible brane is not a local but in fact a global fine-tuning problem. 
Nevertheless, the necessary degree of fine-tuning on the hidden brane is the same as the usual 4D cosmological constant problem, 
as the natural mass scale on the hidden brane (for canonically normalized fields) is given by $\rho_\pi/\omega^2$.  

The effective Planck mass during such a ``remote'' inflation caused by hidden-brane dynamics is given by 
\begin{equation}
M_{\mathrm{Pl},\mathrm{eff}}^2~=~\left.M_{\mathrm{Pl},\mathrm{eff}}^2\right|^{\mathrm{RS1}}_{H=0}\times\left[1+\frac{H^2}{\mu^2}\left(1+\omega-\frac{\omega}{1-\omega}\,2\,\pi\,R\,\mu\right)\right]\;.
\end{equation}
For this scenario the Planck mass is enhanced during inflation, implying that the strength of gravity is reduced. 
Gravity is weakly coupled throughout, meaning that there are no constraints on the possible values of $H$.

Despite the fact that the expansion law looks standard in terms of the canonical hidden-brane energy density, we 
stress that it may not be possible here to directly interpret the effect of the altered effective Planck mass on the tensor-to-scalar ratio.
In particular, inflation is driven by an energy density located on the hidden brane which sharply contradicts 
the IRB assumption that the inflaton dynamics should be confined to the visible brane. 
The results of \cite{Giudice:2002vh} do not simply generalize to cases that violate this assumption. 
A dedicated study would be required to see how density and tensor mode perturbations on the visible 
brane can be affected or even seeded in other cases. 
Given that there is no direct coupling of the inflaton sector to the visible sector, reheating 
could occur via gravitational particle production \cite{Ford:1986sy} (c.f.\ also \cite{Grishchuk:1990bj,Damour:1995pd,Giovannini:1998bp}).
The low efficiency of this reheating mechanism requires the inflation scale to be rather high, definitely well above the BBN scale.
This is no problem here because $H$ is not bounded as discussed above.
Exploring the observational consequences of such a ``remote'' inflation 
scenario is beyond the scope of this work. 

\paragraph{Nihei-Kaloper-Kim-Kim special case.}
So far, we have not specified the mechanism which stabilizes the size of the fifth dimension. 
One possibility to obtain a fixed size of the extra dimension $R$ without any bulk dynamics is by fine-tuning the brane tensions against each other \cite{Nihei:1999mt, Kaloper:1999sm, Kim:1999ja}.
This is a very specific variant of the expanding RS1 case, in the sense that the four-dimensional slices may expand but
the UV and IR brane tensions are fixed relative to each other in order to warrant that the fifth dimension is static. 

With an empty bulk, the general solutions \eqref{eq:00GenSol} receive an additional constraint from the $(55)$ Einstein equation, which is then given by 
\begin{equation}
 4\,f'^2-4\,H^2-\mu^2\,f^2~=~0\;.
\end{equation}
This restricts the coefficients of the general solution to the form
\begin{equation}\label{eq:KimKimCoefficients}
 A~=~\frac{H^2}{\mu^2}\,\widetilde{c_0}\;, \quad\text{and}\quad B~=~\frac{H^2}{\mu^2}\,\frac{1}{\widetilde{c_0}}\;.
\end{equation}
Using the parametrization $\widetilde{c_0}~\equiv~\mathrm{e}^{2\,c_0}$ the metric warping results as
\begin{equation}\label{eq:KimKimSolution}
 f^2(z)~=~\frac{4\,H^2}{\mu^2}\,\sinh^2\left(-\frac{\mu}{2}\,z+c_0\right)\;.
\end{equation}
This is in full agreement with \cite{Kim:1999ja}. Due to the additional constraint, there is one less parameter than in the general solution.
Furthermore, normalizing $f(0)=1$ fixes 
\begin{equation}\label{eq:KKNormalization}
c_0~=~\mathrm{arcsinh}\left(\pm\frac{\mu}{2\,H}\right)\;,
\end{equation}
and there is no free parameter left.

\begin{figure}
\centering
 \includegraphics[width=0.5\textwidth]{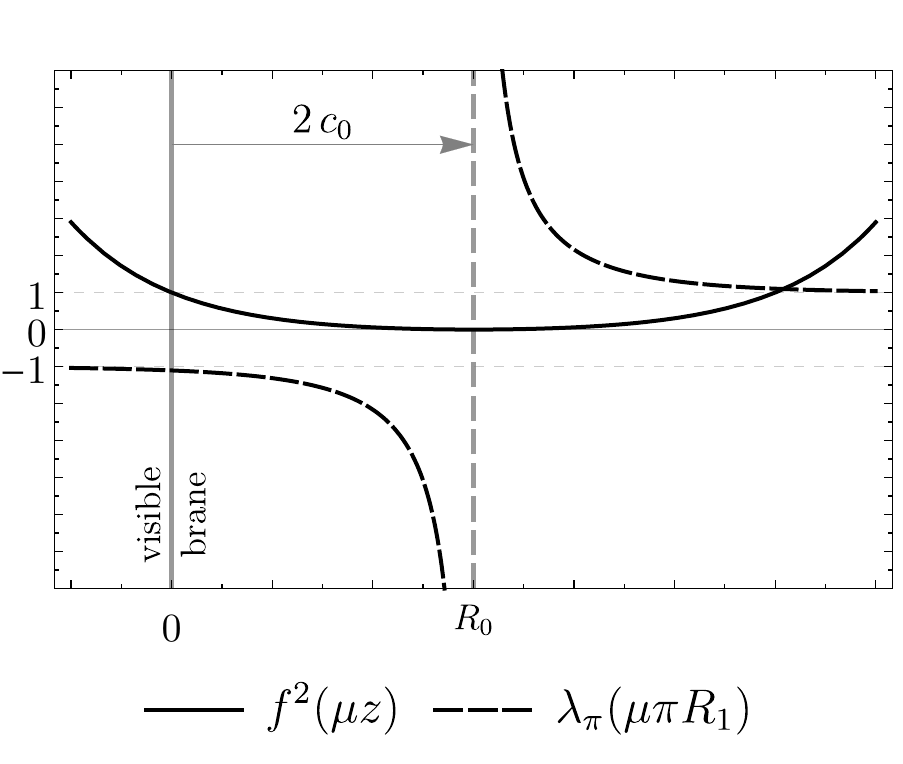}
\caption{The metric warping \eqref{eq:KimKimSolution} of the Nihei-Kaloper-Kim-Kim special case (solid) normalized such that $f(0)=1$. 
The dashed line shows the normalized brane tension $\lambda_\pi(R_1)$ which the hidden brane must carry if it would be located at a distance $R_1$ 
away from the visible brane.  There is a special radius $z=\pi R_{0}$ for which the metric 
is zero and $f'/f$ does not exist. If a brane is placed
at this special radius it decouples and can have any brane tension.}
\label{Fig:KKMetric}
\end{figure}

\begin{figure}
\centering
 \includegraphics[width=0.5\textwidth]{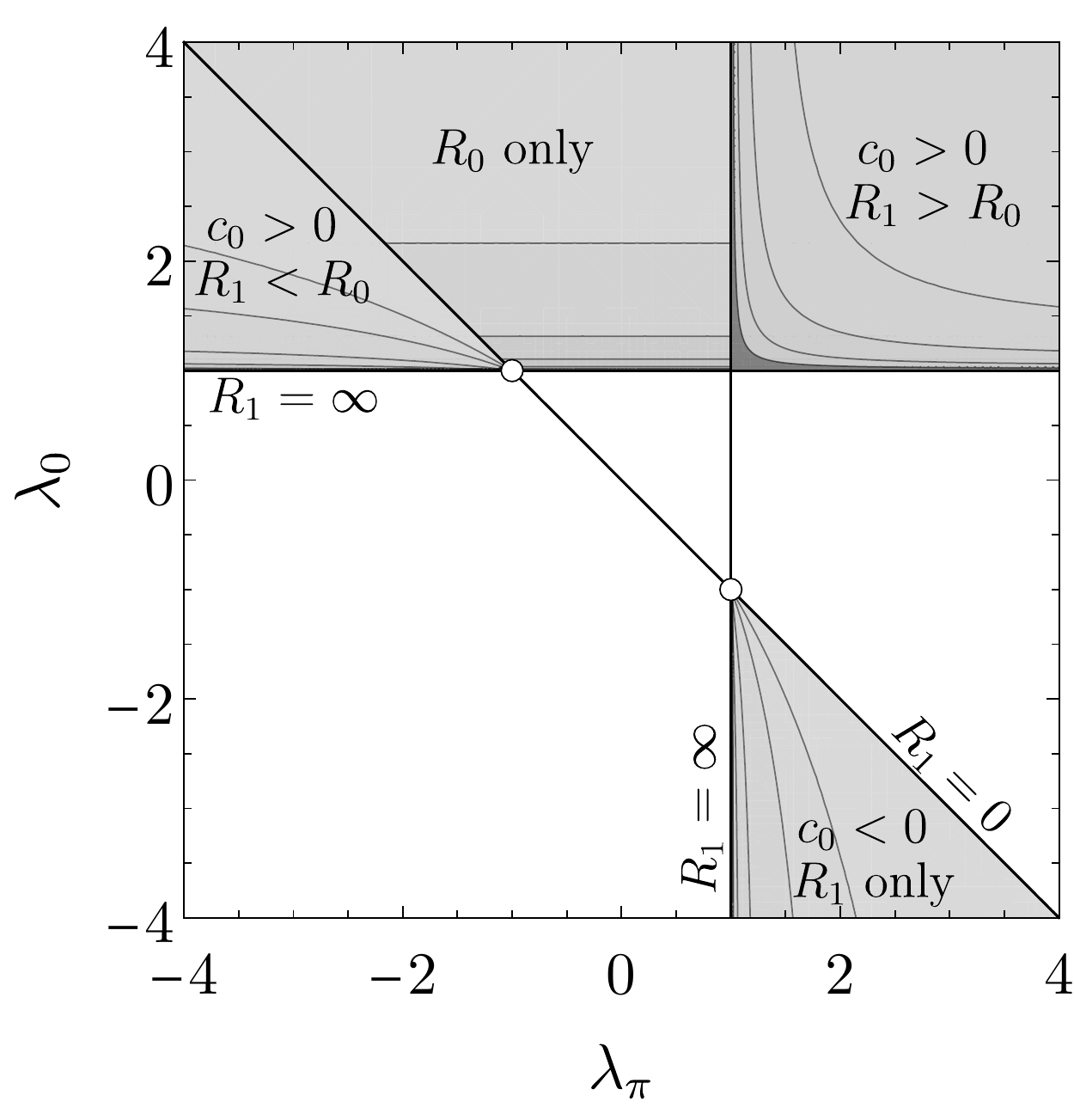}
\caption{Allowed regions (filled) for the normalized brane tensions $\lambda_0,\lambda_\pi$ 
for the Nihei-Kaloper-Kim-Kim special case. The contours illustrate the value of $R_1$ where it is allowed and otherwise the value of $R_{0}$ 
(darker means bigger) such that a consistent solution is achieved. The Hubble rate of the visible brane is everywhere given by \eqref{eq:HKK}.
Contour lines with constant $R$ can also be viewed as parametric curves on which the general expression for $H^2$, \Eqref{eq:ExpansionRate}, 
is reconciled with \eqref{eq:HKK}.
}
\label{Fig:KK}
\end{figure}

The novel constraint is also manifest in the boundary conditions \eqref{eq:RSJumpConditions} which take the form
\begin{subequations}\label{eq:KKBCs}
\begin{align}
  \lambda_0~&=~\coth\,c_0~=~\pm\sqrt{1+\frac{4\,H^2}{\mu^2}}\;,\\
 \lambda_\pi~&=~-\coth\,\left(-\frac{\mu\,\pi\,R}{2}+c_0\right)\;.
\end{align}
\end{subequations}
It immediately follows from eqs.\ \eqref{eq:KKBCs} that $|\lambda_{0,\pi}|>1$ and there is a relation between the size $R$ of the fifth dimension and the brane tensions\footnote{%
As a curiosity, note that this is precisely $2R_\mathrm{RS}$ \eqref{eq:SaticRadius}.}
\begin{equation} \label{R1}
 R~=~R_1~\equiv~\frac{1}{\mu\,\pi}\,\ln\left[\frac{(\lambda_0+1)(\lambda_\pi+1)}{(\lambda_0-1)(\lambda_\pi-1)}\right]\;.
\end{equation}
The metric warping together with the boundary condition $\lambda_\pi$ as a function of the chosen radius for a fixed value of $\lambda_0$ is displayed in \Figref{Fig:KKMetric}.
Furthermore, the Hubble rate of the visible brane (as always in our convention at $z=0$) comes out as 
\begin{equation}\label{eq:HKK}
 H^2~=~\frac{\mu^2}{4}\left(\lambda_0^2-1\right)\;,
\end{equation}
while the Hubble rate at the hidden brane (located at $z=\pi R_1$) is given by
\begin{equation} \label{eq:HKK_pi}
 H^2_\pi~\equiv~\frac{H^2}{f_\pi^2}~=~\frac{\mu^2}{4}\left(\lambda_\pi^2-1\right)\;.
\end{equation}
Our solutions fully agree with eqs.\ (16) and (20) of \cite{Kim:1999ja} 
after noting that their $k\equiv\mu/2$, $L_5\equiv \pi\,R$, and $k_{1,2}/k\equiv\lambda_{0,\pi}$.
However, the discussion in \cite{Kim:1999ja} 
was limited to the parameter region $1<\lambda_0<-\lambda_\pi$.
This limitation makes sense if the hierarchy problem is to be solved at the $z=\pi R_1$ brane and if
one requires that the metric should not have a zero at any point in the extra dimension.
On the other hand, if we do not impose these requirements we find that there are additional regions in parameter 
space for which a consistent solution can be found, c.f.\ \Figref{Fig:KK} (see also \cite{Nihei:1999mt, Kaloper:1999sm}).
In particular, we find that there is another solution for the size of the extra dimension 
\begin{equation}
 R~=~R_{0}~\equiv~\frac{1}{\mu\,\pi}\,\ln\left[\frac{(\lambda_0+1)}{(\lambda_0-1)}\right]\;,
\end{equation}
for which the general solution for $H^2$ \eqref{eq:ExpansionRate} is reconciled with \eqref{eq:HKK}.
This solution makes sense when $\lambda_0>1$ (i.e.\ $c_0>0$), irrespective of the value of $\lambda_\pi$.
For this solution, the fine-tuned value of the radius is independent of the energy density on the hidden brane! 
In order to understand why this is the case, it is important to note that $z=\pi R_{0}$ corresponds to the zero 
of the metric \eqref{eq:KimKimSolution} (cf.\ \Figref{Fig:KKMetric}). Despite the fact that $f'(\pi R_{0})/f(\pi R_{0})$ 
does not exist, this second solution for the radius is consistent with the boundary conditions 
for any value of $\lambda_\pi$. This may appear surprising in view of the condition \eqref{eq:KKBCs}.
However we note that in deriving \eqref{eq:JumpCondition}, which is the general origin of \eqref{eq:KKBCs},  
it was tacitly assumed that the metric has no zero $f(z_0)\neq0$. 
Without imposing this requirement the actual boundary conditions for a brane at $z=z_0$ read
\begin{equation}
 -3\,f(z_0)\left[f'(z_0^{+})-f'(z_0^{-})\right]~=~f(z_0)^2\,\kappa^2\,\rho_{z_0}\;.
\end{equation}
It follows that at metric zeros $f(z_0)=0$ the boundary conditions are trivially fulfilled 
and the parameter $\rho_{z_0}$ will not enter the solution. 
The expansion rate of the hidden brane, $H_{z_0}$, becomes meaningless in this case.
We conclude that putting the hidden brane precisely at the zero of the metric is a valid solution
for which the hidden brane decouples and does not play any role for physics on the visible brane.

Let us discuss how the hierarchy 
problem can be addressed in the NKKK setup, while pointing out the effect on the inflationary tensor modes.
The (orbifolded) extra dimension extends between the visible brane at $z=0$ and a hidden brane at $z=\pi R_1$ with
the general solution for the metric given by \eqref{eq:KimKimSolution}.
Integrating out the extra dimension, we find the strength of effective four-dimensional gravity to be 
\begin{equation}\label{eq:PlEffKandK}
 M_{\mathrm{Pl},\mathrm{eff}}^2~=~\left.M_{\mathrm{Pl},\mathrm{eff}}^2\right|^{\mathrm{RS1}}_{H=0}\,\times\,\frac{H^2}{\mu^2}\,\left[\e^{\pm2\,\arcsinh\,\frac{\mu}{2\,H}}+\omega\,\e^{\mp2\,\arcsinh\,\frac{\mu}{2\,H}}-\frac{\omega}{1-\omega}\,2\,\pi\,R\,\mu\right]\;.
\end{equation}
The two signs correspond to the two cases $c_0\lessgtr0$. There are no restrictions on the possible values of $H$ and $R$ from this since $M_{\mathrm{Pl},\mathrm{eff}}^2$ is strictly positive.
In all cases where the hierarchy problem is solved, the Planck mass is enhanced during inflation implying that gravity is weakened and the tensor mode perturbations are reduced.
For $c_0\lessgtr0$ and taking the limit $H\rightarrow0$ (corresponding to $|\lambda_0|\rightarrow1$) the metric approaches the standard RS warped 
form $f(z)^2=\e^{\pm\mu z}$ with an effective Planck mass
\renewcommand{\arraystretch}{1.45}
\begin{align}
 M_{\mathrm{Pl},\mathrm{eff}}^2~\xrightarrow{H\rightarrow0}~\left\{\begin{array}{lll} \displaystyle\left.M_{\mathrm{Pl},\mathrm{eff}}^2\right|^{\mathrm{RS1}}_{H=0}\;, & \mathrm{for} & c_0~<~0\;, \\
 \displaystyle\left.M_{\mathrm{Pl},\mathrm{eff}}^2\right|^{\mathrm{RS2}}_{H=0}~\equiv~\frac{2\,M^3}{\mu}\left(1-\omega\right)\;, & \mathrm{for} & c_0~>~0\;. \\ \end{array}\right.
\end{align}
The hierarchy problem at the visible brane (without loss of generality placed at $z=0$) can be reliably addressed only in the case $c_0<0$
where the metric warping increases away from the brane.

In the case $c_0>0$ and $R_1> R_0$ 
the zero of $f(z)^2$ is present in the bulk. 
In contrast to the cases discussed in the previous chapters, 
the zero of $f(z)^2$ does not correspond to a true (e.g.\ curvature) singularity but
is instead related to the presence of a causal horizon in the extra dimension \cite{Kaloper:1999sm,Langlois:2000ns}.
Following, for example, \cite{Langlois:2000ns,Gomez:2000bu} one can restrict the size of the extra dimension to a causally connected region, 
i.e.\ let the extra dimension end on the horizon at $\pi R_0=2c_0/\mu$.
In this case the integral \eqref{eq:EffectivePlanckMass} results in
\begin{equation}\label{eq:PlEffKandKHorizon}
 M_{\mathrm{Pl},\mathrm{eff}}^2~=~
 \frac{2\,M^3}{\mu}\,\frac{2\,H^2}{\mu^2}\left[\sinh\left(2\,\arcsinh\frac{\mu}{2\,H}\right)-2\,\arcsinh \frac{\mu}{2\,H}\right]\;.
\end{equation}
This case cannot address the hierarchy problem. The effective Planck mass only implicitly depends on the size of the extra 
dimension via $H$ which is in a one-to-one relationship with $R$. 
The extra dimension can become infinite in size only if $H\rightarrow0$, in which case we approach the RS2 limit.

A naive second possibility to address the hierarchy problem would be $c_0>0$ with $R_1>R_0$, 
corresponding to the upper right region of \Figref{Fig:KK} where $\lambda_0, \lambda_\pi > 1$ 
and the metric is cusped upwards on both branes (cf.\ \Figref{Fig:KKMetric}).
However, in this case the causal horizon is present in the bulk and the extra dimension consists of two causally disconnected regions without interaction. 
The graviton zero mode then would have to be normalized in causally connected regions only, 
corresponding to the previously discussed case that the extra dimension ends on the horizon \eqref{eq:PlEffKandKHorizon}.

A different possibility could arise if the NKKK case is only the late time limit 
to a situation in which there was initially no horizon but it formed dynamically.
Just before the time of horizon formation the graviton zero mode would have to be normalized to the full
extra dimension and it remains to be investigated in a fully dynamical setting how this normalization 
would change after the horizon is formed.
Exploring this possibility would require a fully dynamical treatment of the extra dimension
and the formation of the horizon, which is beyond the scope of this work and so we will not further 
discuss this here.

\paragraph{Expanding RS2.}
Let us also discuss the RS2 model \cite{Randall:1999vf}. Here, the brane with positive tension is the visible brane centered at $z=0$, 
and the radius of the extra dimension is eventually taken to infinity. 
This model does not solve the hierarchy problem, but it is 
of interest simply due to the fact that an infinite extra dimension is allowed in consistency with observation.

The original (static) RS2 setup is obtained by choosing $\lambda_0=-\lambda_\pi=1$. In this case the effective four-dimensional
Planck mass is
\begin{equation}\label{eq:MPlstaticRS2}
 \left.M^2_\mathrm{Pl,eff}\right|^{\mathrm{RS2}}_{H=0}~=~\frac{2\,M^3}{\mu}\left(1-\omega\right)~=~\frac{2\,M^3}{\mu}\left(1-\mathrm{e}^{-\mu\,\pi\,R}\right)\;.
\end{equation}
There is no exponential hierarchy generated between the 5D and 4D scales, and there is no obstruction in taking $R\rightarrow\infty$.

For the expanding case, we assume that the inflaton sector is confined to the visible brane. Therefore, we 
keep $\lambda_\pi=-1$ but add a surplus energy density to the visible brane given by $\lambda_0=1+\epsilon_0$, with
$\epsilon_0$ as given in \eqref{eq:epsilonDef}. The general results \eqref{eq:00GenSol} and \eqref{eq:ExpansionRate} apply and one finds a Hubble rate
\begin{equation}
 H^2~=~\frac{\mu^2\,\epsilon_0}{2(1-\omega)}~=~\frac{\rho_0}{3\,\displaystyle{\left.M^2_\mathrm{Pl,eff}\right|^{\mathrm{RS2}}_{H=0}}}\;,
\end{equation}
which is consistent with the usual four-dimensional expansion law.

The metric warping is given by
\begin{equation}
 f^2(z)~=~\left[ 1+\frac{H^2}{\mu^2}\left(2-\omega\right)\right]\e^{-\mu\,z}+\frac{H^2}{\mu^2}\,\omega\,\e^{\mu\,z}-\frac{2\,H^2}{\mu^2}\;,
\end{equation}
and the resulting effective Planck mass is 
\begin{equation}
 M_{\mathrm{Pl},\mathrm{eff}}^2~=~\left.M_{\mathrm{Pl},\mathrm{eff}}^2\right|^{\mathrm{RS2}}_{H=0}\times\left[1-\frac{H^2}{\mu^2}\left(\omega-3+\frac{2\,\mu\,\pi\,R}{1-\omega}\right)\right]\;.
\end{equation}
The correction factor always reduces the effective Planck mass, i.e.\ it increases the strength of gravity during inflation.
Since the discussion of \cite{Giudice:2002vh} is fully applicable, we conclude that the tensor modes during inflation are enhanced in the RS2 setup.

Analogous to the LED and RS1 cases the metric can cross zero at which point a curvature singularity appears in the bulk. Avoiding this situation imposes the bound
\begin{equation}
H^2~<~\mu^2\,\omega\;,
\end{equation}
which also ensures $M_{\mathrm{Pl},\mathrm{eff}}^2>0$.
Obeying this bound enforces $H\rightarrow0$ as $R\rightarrow\infty$ implying that the inflating RS2 case is inconsistent with taking the size of the extra dimension to infinity.
In principle this is nothing new, as it also happens in the LED case, cf.\ \Eqref{eq:EffectivePlanckMassLED} and the related discussion in \cite{Giudice:2002vh}.
The fact that we cannot take the size of the extra dimension to infinity without letting $H\to0$
is independent of whether we take into account the (55) equation or not, as is clear from the previous section.
This conclusion does not change in case one allows for other values of $\lambda_\pi$ (for example, $\lambda_\pi=0$ or $\lambda_\pi=-1+\epsilon_0$).         

\section{Linear Dilaton Model}
\label{sec:LD}
\subsection{Generalities}

Let us consider the linear dilaton configurations \cite{Aharony:1998ub,Antoniadis:2001sw,Antoniadis:2011qw,Cox:2012ee,Baryakhtar:2012wj} 
of little string theory (see \cite{Aharony:1999ks} for more references on LST).
This case is akin to the RS case in the sense that there will be a negative bulk cosmological constant and two branes. The crucial difference to the RS case is
the presence of an additional scalar field, the dilaton. Ultimately, this gives rise to power law warping in contrast to the exponential warping of the RS case.

The action in the Einstein frame is given by \cite{Antoniadis:2011qw,Cox:2012ee,Giudice:2016yja,Kehagias:2016kzt}
\begin{equation}\label{eq:CWaction}
\begin{split}
 S~=~\int\mathrm{d}^4x\,\mathrm{d}z\,\sqrt{|g|}\left\{ \frac{M^3}{2} \right.&\left.\left(- \mathcal{R} + \frac{1}{3}\,g^{MN}\,\partial_M S\,\partial_N S+4\,k^2\,\mathrm{e}^{-\frac{2}{3}\,S}\right)\right. \\
 +& \left. \frac{\mathrm{e}^{-\frac{1}{3}\,S}}{\sqrt{|g_{55}|}}\left[\mathcal{L}_0\,\delta(z)+\mathcal{L}_\pi\,\delta(z-\pi\,R)\right]+\mathcal{L}_b(x,z)\right\}\;,
\end{split}
\end{equation}
where $-2k^2\equiv\kappa^2\Lambda$ and we have allowed for the possibility of having branes and extra contributions in the bulk.
$S$ is the (dimensionless) dilaton and the corresponding canonically normalized field is $\phi=M^{3/2}S/\sqrt{3}$.

The dilaton field itself can be used to stabilize the radius of the fifth dimension.
Alternatively, one could also introduce a Goldberger-Wise (GW) type scalar field \cite{Goldberger:1999uk} fulfilling the same purpose. 
In analogy to the LED and RS case, we will first be agnostic about the details of the
radius stabilization and simply ignore the corresponding $(55)$ Einstein equation, assuming that it
gives rise to a stable radius. We are again looking for solutions to Einstein's equations that satisfy the metric ansatz
\begin{equation}\label{eq:CWansatzMetric}
 \mathrm{d}s^2~=~f(z)^2\,\left(\mathrm{d}t^2-a(t)^2\,\delta_{ij}\,\mathrm{d}x^i\mathrm{d}x^j\right)-\mathrm{d}z^2\;.
\end{equation}
The Einstein tensor is the same as above, while the energy-momentum tensor now is given by
\begin{equation}
 \kappa^2\,T_{MN}~=~\frac13\,\partial_M S\,\partial_N S-\frac16\,g^{PQ}\,\partial_P S\,\partial_Q S\,g_{MN}+\kappa^2\,\Lambda\,\mathrm{e}^{-2\,S/3}\,g_{MN}\;.
 \end{equation}
The $(00)$ equation in the bulk then can be written as  
\begin{equation}\label{eq:CW00inBulk}
\left(f^2\right)''-2\,H^2+\frac{2}{3}\kappa^2\,\Lambda\,f^2\,\mathrm{e}^{-2\,S/3}+\frac{f^2\,S'^2}{9}~=~0\;.
\end{equation}
Here we are assuming that the dilaton is homogeneous in four dimensions $S=S(z)$. It follows that $H=const.$ and the $(ij)$ and $(00)$ equations are degenerate.
Additionally, there appears the equation of motion of the dilaton, which is in the bulk given by
\begin{equation}\label{eq:DilatonEOM}
 S''+4\,\frac{f'}{f}\,S'+2\,\kappa^2\,\Lambda\,\mathrm{e}^{-2\,S/3}~=~0\;.
\end{equation}
The Bianchi identity is identical to the dilaton equation of motion and does not give an independent constraint.

Due to the exponential dilaton factors in the action, the boundary conditions are modified in comparison to the previous cases.
For a brane at position $z=z_0$ carrying a constant four-dimensional energy density $\mathcal{L}_{z_0}=-\rho_{z_0}$ the
discontinuities of $f'$ and $S'$ across the brane have to fulfill
\begin{equation}\label{eq:CWJumpCondition}
 \frac{f'(z_0^+)-f'(z_0^-)}{f(z_0)}~=~-\frac{\kappa^2\,\rho_{z_0}}{3}\,\mathrm{e}^{-\frac{1}{3}\,S(z_0)}\;,
\end{equation}
as well as
\begin{equation}\label{eq:CWJumpConditionS}
 S'(z_0^+)-S'(z_0^-)~=~-\kappa^2\,\rho_{z_0}\,\mathrm{e}^{-\frac{1}{3}\,S(z_0)}\;.
\end{equation}
We were not able to find a closed form solution for \eqref{eq:CW00inBulk} and \eqref{eq:DilatonEOM} for a general $H\neq0$.
We will, thus, discuss the exactly solved static case $(H=0)$ first, after which we present a perturbative solution 
for the expanding case.

It should be mentioned here that a closed form solution for the LD model in the Jordan frame has been presented in \cite{Meissner:2003qv}.
Despite the difficulties in transforming the solution to the Einstein frame \cite{Meissner:2003qv} their derivation makes use of
the unperturbed (55) equation. This is why we cannot directly adopt their solution here.
Nevertheless, let us emphasize that it would be eminently useful to have an exact solution also for the LD case, just as for the LED and RS cases above.

\subsection{The static case}
Setting $H=0$, the most general simultaneous solution to \eqref{eq:CW00inBulk} and \eqref{eq:DilatonEOM} \textit{in the bulk} is given by 
\begin{equation}\label{eq:LogarithmicDilaton}
 S(z)~=~3\,\ln\,\frac{f(z)}{c_S}\;, \qquad f(z)~=~c_0+\zeta\,c_S\,\frac{2\,k}{3}\,z\;.
\end{equation}
Here, $c_0$ and $c_S$ are arbitrary dimensionless constants and $\zeta=\pm1$ is an undetermined sign.
By rescaling of the four-dimensional coordinates it is always possible to normalize $f(0)=1$, thus, fixing the constant $c_0=1$.
By contrast, $c_S$ corresponds to the normalization of $S(0)$, whose value, however, can also be chosen without loss of generality. 
This can be understood by noting that $c_S$ can be absorbed into $\tilde{k}:=c_S k$ and it thereby disappears completely from 
the bulk action which now contains $\tilde{k}$ and $\tilde{S}(z)=3\ln(1+2\tilde{k}z/3)$ which is automatically normalized to $\tilde{S}(0)=0$. 
Reformulating the brane Lagrangians in terms of $\tilde{S}$ one finds that they have to be globally rescaled by $c_S$, for example \mbox{$\mathcal{L}_0\to c_S\mathcal{L}_0$}. 
To maintain canonically normalized kinetic terms on the branes one then has to rescale the fields in $\mathcal{L}_0$ which likewise leads to an unphysical rescaling of couplings. 
Thereby $c_S$ can be completely absorbed from the theory without loss of generality. 
Correspondingly, a normalization of $S(0)=0$ can always be chosen without physical consequences \cite{Giudice:2016yja}.
We stress this point here because in the expanding ($H\neq0$) case below this conclusion will not hold 
and the physical results change if the boundary condition $S(0)=0$ is changed.
The bulk solution of the static case is consistent with the boundary conditions on the two branes \eqref{eq:CWJumpCondition}, \eqref{eq:CWJumpConditionS} only for the fine-tuned brane tensions\footnote{%
Note that this corresponds to values $-\lambda_0=\lambda_\pi=2/\sqrt{3}$ in our above notation for the RS case, showing that the fine-tuned brane tensions in the CW case 
are different from the ones required in the RS case.}
\begin{equation}
 -\rho_0~=~\rho_\pi~=~\zeta\,4\,k\,M^3\;.
\end{equation}
Here, it makes sense to define the dimensionless quantities 
\begin{equation}
 \eta_0~:=~\frac{\rho_0}{4\,k\,M^3}
 \;,\qquad\text{and}\qquad\eta_\pi~:=~\frac{\rho_\pi}{4\,k\,M^3}\;.
\end{equation}
Choosing $\zeta=+1$, i.e.\ $\eta_0=-\eta_\pi=-1$ and assuming the usual orbifold symmetry $z\to-z$ for solutions in the two separate domains, one obtains 
\begin{equation}\label{eq:staticSols}
f(z)~=~f_s(z)~:=~1+\frac{2\,k}{3}\,|z|\;\quad\text{and}\qquad S(z)~=~S_s(z)~:=~3\,\ln\,\left(1+\frac{2\,k}{3}\,|z|\right)\;.
\end{equation}
This is the standard linear dilaton solution (see e.g.\ \cite{Antoniadis:2011qw,Cox:2012ee,Baryakhtar:2012wj}) which appears here as ``logarithmic dilaton'' 
due to our euclidean coordinate choice for the fifth dimension.
Compared to the exponential warping in the static RS metric \eqref{eq:RSMetric} we find here a power-law warping. Taking the 
fifth dimension to be of size $z\in\left[-\pi R,\pi R\right]$ the effective four-dimensional Planck mass is given by
\begin{equation}
 \left.M_{\mathrm{Pl},\mathrm{eff}}^2\right|^\mathrm{LD}_{H=0}~=~2\,M^3\,\int_0^{\pi\,R}\,\mathrm{d}z\,f(z)^2~=~\frac{M^3}{k}\left[\left(1+\frac{2\,k}{3}\,\pi\,R\right)^3-1\right]\;.
\end{equation}
Taking the fundamental scale to be $M\sim k\sim\mathrm{TeV}$ and requiring the observed value for $M_{\mathrm{Pl},\mathrm{eff}}$ we find that $k\pi R\gtrsim 10^{11}$, corresponding to an extra dimension of size $\sim 10\,\mathrm{nm}$. 

Note that it is crucial to choose the same sign for $\zeta$ and the possible values of $z$. Choosing $\zeta=-1$ (or equivalently allowing for negative values of $z$)
the metric vanishes and the dilaton profile diverges at $z_\mathrm{sing}=(-)3/2k$ corresponding to a physical singularity. This would give rise to a natural cutoff size of 
the extra dimension $\pi R\leq3/2k$. The presence of such singularities has already been noted in \cite{Meissner:2003qv}. 
We are interested in cases where the presence of the extra dimension solves the hierarchy problem. Therefore, we limit ourselves to parameters which allow for an arbitrary size of the extra dimension and avoid
the singularity in the dilaton profile.

\subsection{The expanding case with external stabilization}

Let us now generalize the solution to the case $H\neq0$. Since we were not able to find an exact solution for $f(z)$ and $S(z)$ for the general case,
we will assume that the dimensionless quantity $\delta:=H^2/k^2$ is small and find a perturbative solution in $\delta$. 
In the limit $\delta\rightarrow0$ the linear dilaton solution should be recovered. Therefore, we adopt the ansatz
\begin{equation}\label{eq:LDAnsatz}
 f(z)~=~f_s(z)\left[1+\delta\,df(z)\right]\;,\qquad\text{and}\qquad S(z)~=~S_s(z)+\delta\,dS(z)\;,
\end{equation}
where $f_s(z)$ and $S_s(z)$ are the solutions of the static case given in \eqref{eq:staticSols}. Plugging the ansatz into equations
\eqref{eq:CW00inBulk} and \eqref{eq:DilatonEOM} we expand in $\delta\ll1$ and use that $f_s$ and $S_s$ are solutions of the static case. 
At linear order in $\delta$ we find that $df(z)$ and $dS(z)$ have to fulfill
\begin{align}
 2\,f_s^2\,\frac{df''}{k^2} + \frac{16}{3}\,f_s\,\frac{df'}{k} + \frac{4}{9}\,f_s\,\frac{dS'}{k}+\frac{8}{9}\,dS - 2~&=~0\;,\\
 f_s\,\frac{dS''}{k^2}+\frac83\,\frac{dS'}{k}+\frac{8}{3}\frac{dS}{f_s}+8\,\frac{df'}{k}~&=~0\;.
\end{align}
These two equations can be decoupled, thereby giving rise to a third oder equation for $dS$ which can be solved.
Subsequently the solution for $dS$ can be used in order to solve also for $df$. The corrections to the bulk solutions are then given by
\begin{subequations}\label{eq:LDSolution}
\begin{align}
 dS(z)~&=~ -\frac92 + \frac{c_1}{\left[3\,f_s(z)\right]^3} + \frac{c_2}{\left[3\,f_s(z)\right]^2}+ \frac{c_3}{3\,f_s(z)}\;,\\
 df(z)~&=~ \frac94\,\ln f_s(z) + \frac{c_1}{162\,f_s(z)^3} + \frac{c_2}{54\,f_s(z)^2} + \frac{c_3}{9\,f_s(z)} + c_4\;,
\end{align}
\end{subequations}
with four arbitrary constants $c_{1-4}$. Again, we normalize $f(0)=1$ and $S(0)=0$ by fixing $c_4$ as well as $c_3$, respectively. 

\begin{figure}[t]
\centerline{\subfigure[]{\CenterObject{\includegraphics[width=0.48\textwidth]{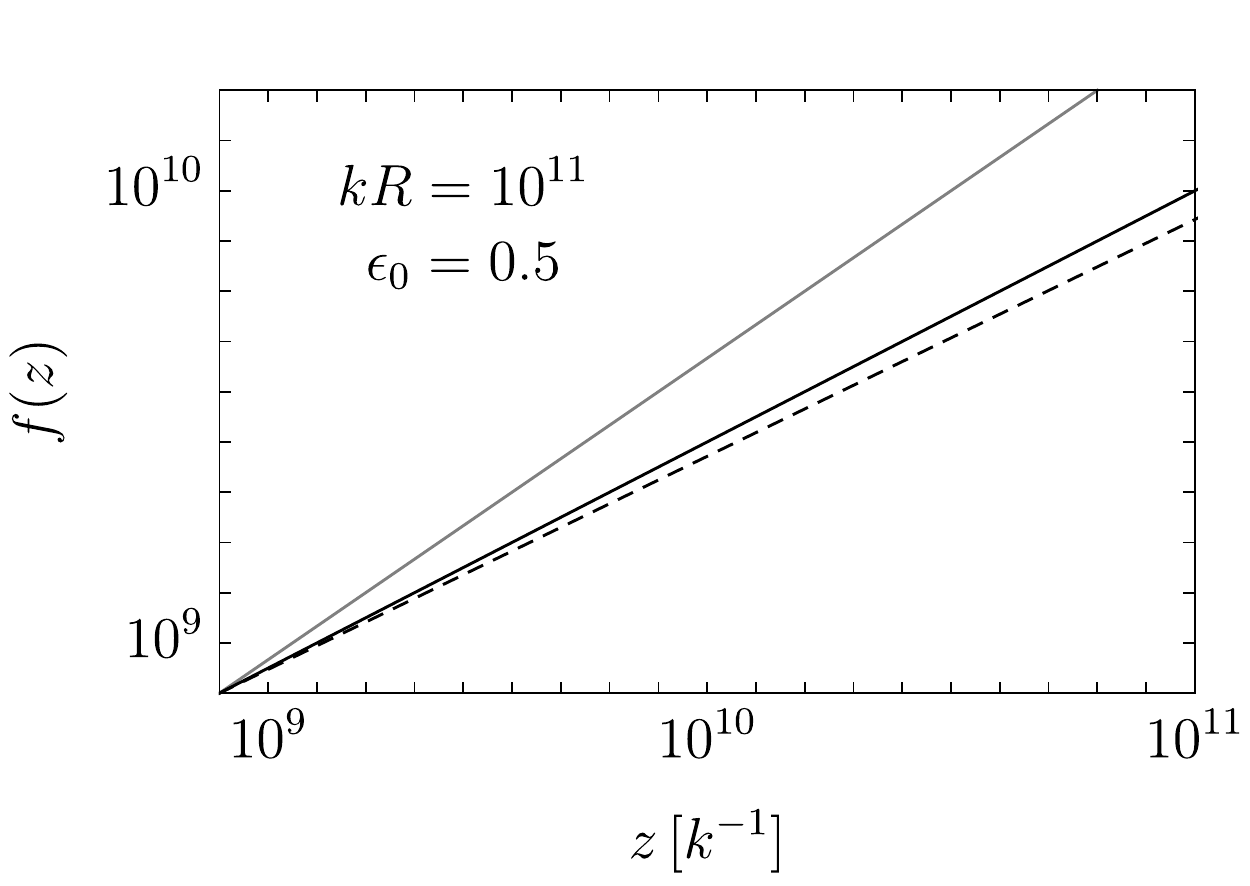}}}
\qquad\subfigure[]{\CenterObject{\includegraphics[width=0.44\textwidth]{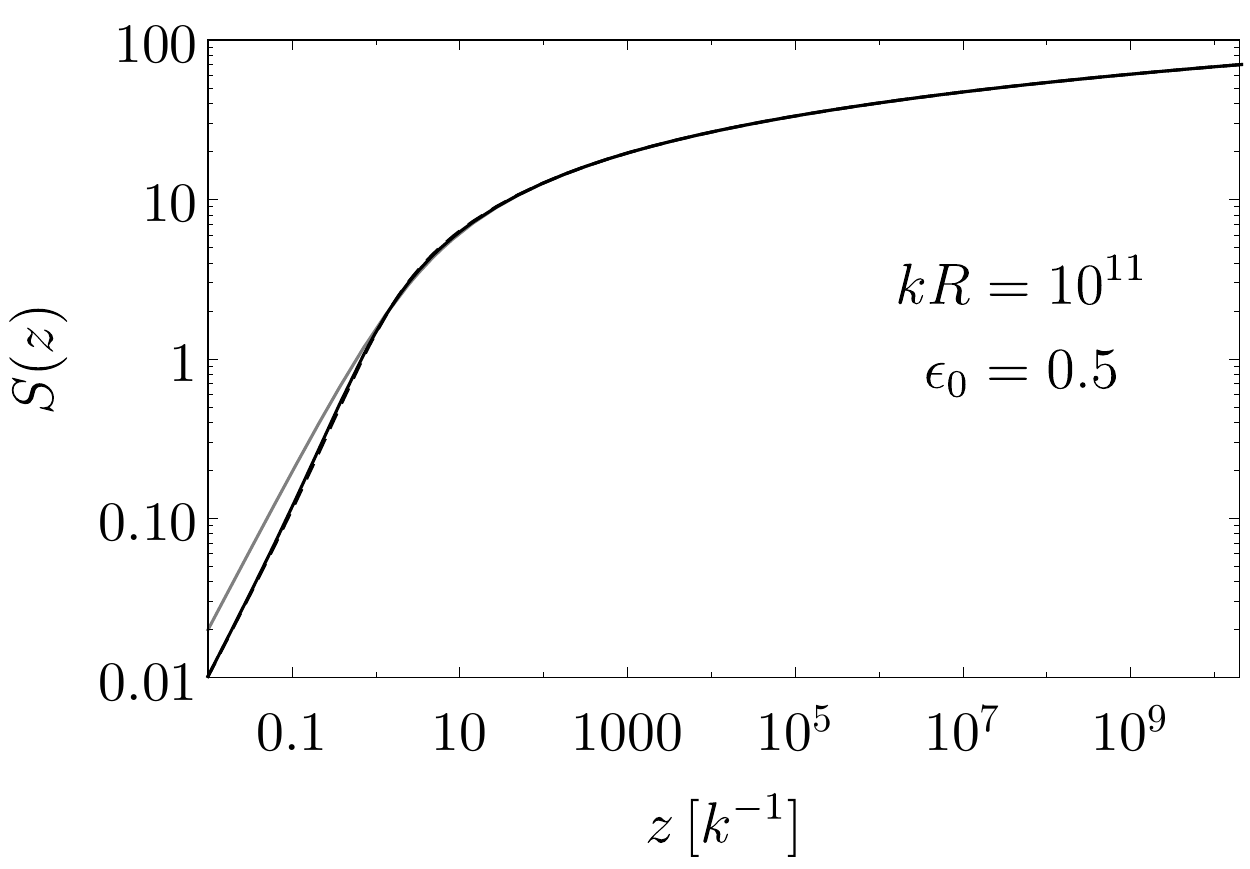}}}}
  \caption{Metric warping (a) and Dilaton profile (b) of the Linear Dilaton model with inflation (black) as compared to the static solution (gray). Next to our perturbative results (black) we also show a numerical solution (black, dashed).
  For the chosen parameters, the resulting Hubble rate is $H\approx4\times10^{-12}k\approx4\times10^{-9}\,\GeV$.}
  \label{Fig:LD}
\end{figure}

The junction conditions on the branes are
\begin{subequations}\label{eq:CWJumpConditions}
\begin{align}
 \frac{f'(0)}{f(0)}~&=~-\frac23\,k\,\eta_0\,\e^{-S(0)/3}\;,& \qquad\frac{f'(\pi R)}{f(\pi R)}~&=~\frac23\,k\,\eta_\pi\,\e^{-S(\pi R)/3}\;,& \\
 S'(0)~&=~-2\,k\,\eta_0\,\e^{-S(0)/3}\;,& \qquad S'(\pi R)~&=~2\,k\,\eta_\pi\,\e^{-S(\pi R)/3}\;.& 
\end{align}
\end{subequations}
As our bulk solution is only valid up to order $\delta$ we can only require that the boundary conditions are solved up to that order.
This implies that deviations of the brane tensions should be small compared to the static case, i.e.\ $\eta_0=-1+\epsilon_0$ and $\eta_\pi=1+\epsilon_\pi$
with $\epsilon_0, \epsilon_\pi\ll 1$. The boundary conditions are then solved by
\begin{align}\nonumber
 c_1~&=~\frac{243}{2}\left(f_{s,\pi}^2+f_{s,\pi}\right)\;,& c_2~&=~-\frac{243}{4}\left(f_{s,\pi}^2+f_{s,\pi}+1\right)\;,& \\ \label{eq:SolLD}
 c_3~&=~\frac{27}{4}\left(f_{s,\pi}^2+f_{s,\pi}+5\right)\;,&
\end{align}
together with the relations 
\begin{equation}\label{eq:CWBoundariesSolved}
\delta~=~\frac{4\,\epsilon_0}{3\left(f_{s,\pi}^2+f_{s,\pi}-1\right)}\;,\qquad\text{and}\qquad \frac{\epsilon_\pi}{\epsilon_0}~=~\frac{f_{s,\pi}^2-f_{s,\pi}-1}{f_{s,\pi}^2 \left(f_{s,\pi}^2+f_{s,\pi}-1\right)}\;.
\end{equation}
Here we have used the abbreviation $f_{s,\pi}\equiv f_{s}(\pi R)$.
Finally, our solution for the metric warping is fully specified and given by
\begin{equation}
 f(z)~=~1+\frac{2\,k}{3}\,|z|-H^2\,\pi^2\,R^2\,k\,|z|\,\frac{2\,\left(9+3\,k\,|z|+2\,k^2\,z^2\right)}{9\,\left(3+2\,k\,|z|\right)^2}~+~\mathrm{h.o.}\;,
\end{equation}
where $\mathrm{h.o.}$ denotes terms of higher order in $H^2/k^2\ll1$ or lower order in $k\pi R\gg 1$.
For completeness, we also state the leading order correction to the dilaton profile which is given by
\begin{equation}
 S(z)~=~3\,\ln\,\left(1+\frac{2\,k}{3}\,|z|\right)+ H^2\,\pi^2\,R^2\,k\,|z|\,\frac{6\,\left(2\,k\,|z|-3\right)}{\left(2\,k\,|z|+3\right)^3}~+~\mathrm{h.o.}\;.
\end{equation}
Our perturbative solutions agree well with a numerical solution of \eqref{eq:CW00inBulk} and \eqref{eq:DilatonEOM}, as displayed in \Figref{Fig:LD}. 

Given the metric warping $f(z)$, the effective Planck mass during inflation can be computed to be 
\begin{equation}\label{eq:PlEffCW}
 M_{\mathrm{Pl},\mathrm{eff}}^2~=~\left.M_{\mathrm{Pl},\mathrm{eff}}^2\right|^{\mathrm{LD}}_{H=0}\times\left(1-\frac13\,H^2\,\pi^2\,R^2 +\mathrm{h.o.} \right)\;.
\end{equation}
The inferred bound on $HR$ from the perturbativity requirement on gravity $M_{\mathrm{Pl},\mathrm{eff}}^2>0$ during inflation is
\begin{equation}
 H^2\,\pi^2\,R^2~\lesssim~3\;,
\end{equation}
corresponding to $H<10\ev$ (i.e.\ $T_\mathrm{RH}<10^5\GeV$) for $R\sim10\,\mathrm{nm}$.
Note that it is not possible here to straightforwardly interpret the decrease of the Planck mass during inflation in terms of the tensor-to-scalar ratio, as we will discuss in the following.

The first relation in \eqref{eq:CWBoundariesSolved} can be written as 
\begin{equation}
 H^2~=~\frac{4\,k^2\,\epsilon_0}{3\left(f_{s,\pi}^2+f_{s,\pi}-1\right)}~\approx~\frac{3\,\epsilon_0}{\pi^2\,R^2}~=~
 \frac{2}{3}\,k\,\pi\,R~\times~\frac{\rho_0}{3\,\displaystyle{\left.M_{\mathrm{Pl},\mathrm{eff}}^2\right|^{\mathrm{LD}}_{H=0}}}\;,
\end{equation}
where $\rho_0$ denotes the surplus energy density on the visible brane and we have expanded in $k\pi R\gg 1$ to simplify the result.
Note that we do \textit{not} recover the standard expansion law on the visible brane unlike in the LED or RS case with inflation. 
Just like in the cases of ``remote'' inflation in the RS model and the special case of Nihei-Kaloper-Kim-Kim,
the expansion law of the visible brane is non-standard. 
The origin of the non-standard expansion law here is the necessary relation between the surplus energy densities on 
the IR and UV branes, manifest in the second equation of \eqref{eq:CWBoundariesSolved}. 
Such a relation is, of course, inconsistent with the picture of having the inflaton sector confined to one of the branes.
Restoring the individual contributions of the two brane energy densities to the expansion one can write 
\begin{equation}
 H^2~\approx~\frac{\rho_0}{3\,\displaystyle{\left.M_{\mathrm{Pl},\mathrm{eff}}^2\right|^{\mathrm{LD}}_{H=0}}}+
 \frac{8}{27}\,k^3\,\pi^3\,R^3\times\frac{\rho_\pi}{3\,\displaystyle{\left.M_{\mathrm{Pl},\mathrm{eff}}^2\right|^{\mathrm{LD}}_{H=0}}}\;.
\end{equation}
This shows that if one would ignore the required interrelation of energy densities and simply set $\rho_\pi\to0$ the
standard expansion law on the visible brane would be recovered. However, such an ad-hoc prescription is inconsistent with our solution, 
in particular with the relation \eqref{eq:CWBoundariesSolved}.
In this sense the common wisdom, that a stabilized radius leads to a standard inflation law on the brane,
does not hold for the Linear Dilaton model.

Clearly, the requirements to apply the analysis of \cite{Giudice:2002vh} are not fulfilled here. 
In particular, it is not possible to assume that the energy density that drives inflation -- i.e.\ the inflaton and its potential -- is confined to the visible brane.
A dedicated study would be required to track the impact of the non-standard expansion law on the observable scalar and tensor mode perturbations after inflation.

Let us remark that the physical origin of the relation between the brane energy densities in (\ref{eq:CWBoundariesSolved}), which leads to the non-standard expansion law, 
is the dilaton degree of freedom. 
In the present case, the dilaton dynamics does not decouple from the system even in low energy limit so that the correlation of the two brane energy densities still holds.
In a sense, the situation is similar to the NKKK case in the RS model. In that case, the radion is massless so that the (55) Einstein equation 
does not decouple from the system. This in turn requires the brane energy density relation (\ref{R1}) for a given radius $R$, which leads to the non-standard expansion law (\ref{eq:HKK}).
Just as the correlation (\ref{R1}) becomes irrelevant when the radion is heavy in the RS model, also the relation (\ref{eq:CWBoundariesSolved}) in the Linear Dilaton model will be broken  
if the dilaton gets massive. We will see this to be the case in the next subsection where we introduce additional brane localized dilaton potentials,
which make the dilaton fluctuations over the background solution heavy enough to decouple them from the system in the low energy limit.
 
So far we have been agnostic about the details of the stabilization mechanism. 
The most economic way to stabilize the extra dimension in the Linear Dilaton model is to invoke the 
dilaton field itself \cite{Cox:2012ee}. 
If $S$ experiences strong boundary potentials, the field values $S_s(0)=S_{0}$ and $S_s(\pi R)=S_\pi$ on the branes are fixed, 
corresponding to two additional boundary conditions. \textit{In the static case}, the size of the extra dimension then 
is determined by the relation \cite{Cox:2012ee}
\begin{equation}\label{eq:StaticRadius}
 k\,\pi\,R~=~\frac32\,\left[\exp\left(\frac{S_\pi-S_0}{3}\right)-1\right]\;,
\end{equation}
completely analogous to the usual Goldberger-Wise mechanism \cite{Goldberger:1999uk}.
An analogous scheme has been adopted for the expanding case in \cite{Kehagias:2016kzt}. 
However, we find that one has to be very careful in applying the stabilization scheme of the static case for the expanding case.
The reason is that the boundary conditions \eqref{eq:CWJumpConditions} are modified by the brane potentials of $S$.
This does not affect the solution of the static case, as the modified boundary conditions are automatically fulfilled.
In the dynamical case, however, the change is important as we will see in the following.

\subsection{The expanding case with dilaton stabilization}
\label{sec:LDS}

Let us consider the case that the dilaton itself is used as a stabilizer. 
The dilaton-stabilized solution is somewhat more elaborate than the stabilization 
by an additional Goldberger-Wise scalar, simply due to the fact that the back reaction of the stabilizing 
field is fully accounted for in the computation of the metric. Assuming an otherwise empty bulk, the bulk Lagrangian 
is fully specified and the (55) Einstein equation should be taken into account. 
In addition, the boundary conditions \eqref{eq:CWJumpConditions} are modified by the brane potentials of $S$.

The (55) equation in the bulk is given by
\begin{equation}
 4\,f'^2-4\,H^2-\frac19\,f^2\,S'^2+\frac23\,\kappa^2\,\Lambda\,f^2\,\e^{-2\,S/3}~=~0\;.
\end{equation}
Stabilization can be achieved by imposing strong boundary potentials for $S$ which can be modeled in 
\eqref{eq:CWaction} by the choice 
\begin{align}
 \mathcal{L}_0~&=~-\rho_0-\xi_0\,M^4\left(S_0-S(z)\right)^2\;,\\
 \mathcal{L}_\pi~&=~-\rho_\pi-\xi_\pi\,M^4\left(S_\pi-S(z)\right)^2\;.
\end{align}
Here, $\xi_{0,\pi}$ are dimensionless parameters that characterize the strength of the respective potential. 

This modifies the boundary conditions \eqref{eq:CWJumpCondition} and \eqref{eq:CWJumpConditionS} at the respective position $z_0=\left\{0,\pi R\right\}$ to
\begin{subequations}
\begin{align}
 \frac{f'(z_0^+)-f'(z_0^-)}{f(z_0)}~&=~-\frac{\kappa^2}{3}\,\mathrm{e}^{-S(z_0)/3}\left[
 \rho_{z_0}+\xi_{z_0}\,M^4\left(S_{z_0}-S(z_0)\right)^2
 \right]\;,\\ \label{eq:CWJumpConditionLDS}
 S'(z_0^+)-S'(z_0^-)~&=~-\kappa^2\,\mathrm{e}^{-S(z_0)/3}\left[
 \rho_{z_0}+\xi_{z_0}\,M^4\left(S_{z_0}-S(z_0)\right)^2+6\,\xi_{z_0}\,M^4\left(S_{z_0}-S(z_0)\right)
 \right]\;.
\end{align}
\end{subequations}
Here, we take the boundary potential parameters $S_{0}$ and $S_{\pi}$ 
to be the same for the static and the dynamic case.

In the static case, the solutions \eqref{eq:staticSols} are compatible with the 
modified boundary conditions if the additional constraints $S_s(0)=S_{0}$ and $S_s(\pi R)=S_\pi$ are fulfilled.
Furthermore, the (55) Einstein equation is automatically fulfilled. 
Thus, the original static case solution is completely consistent also with the assumption of dilaton stabilization and one eventually arrives at \eqref{eq:StaticRadius}
which determines the stabilized size of the extra dimension.

\begin{figure}[t]
\centerline{\subfigure[]{\CenterObject{\includegraphics[width=0.48\textwidth]{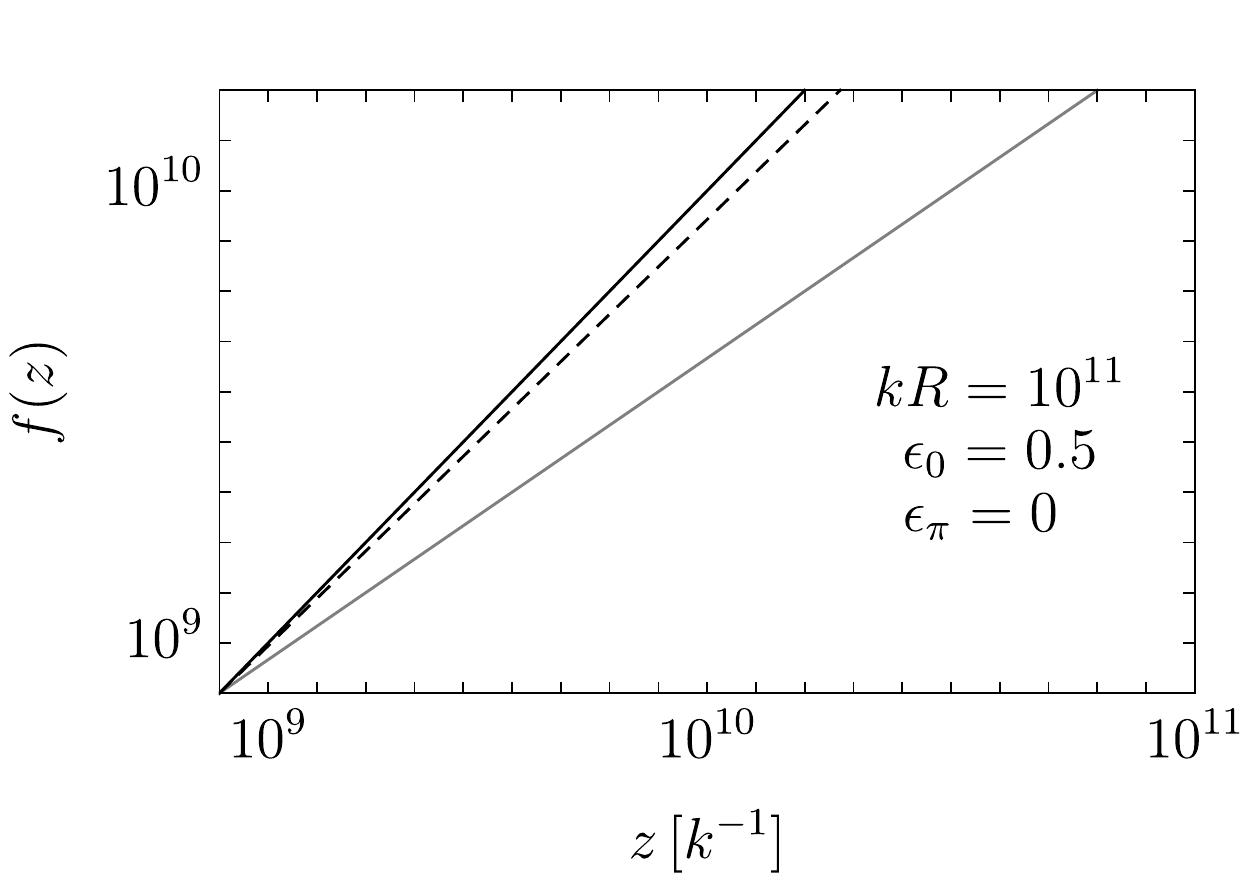}}}
\qquad\subfigure[]{\CenterObject{\includegraphics[width=0.44\textwidth]{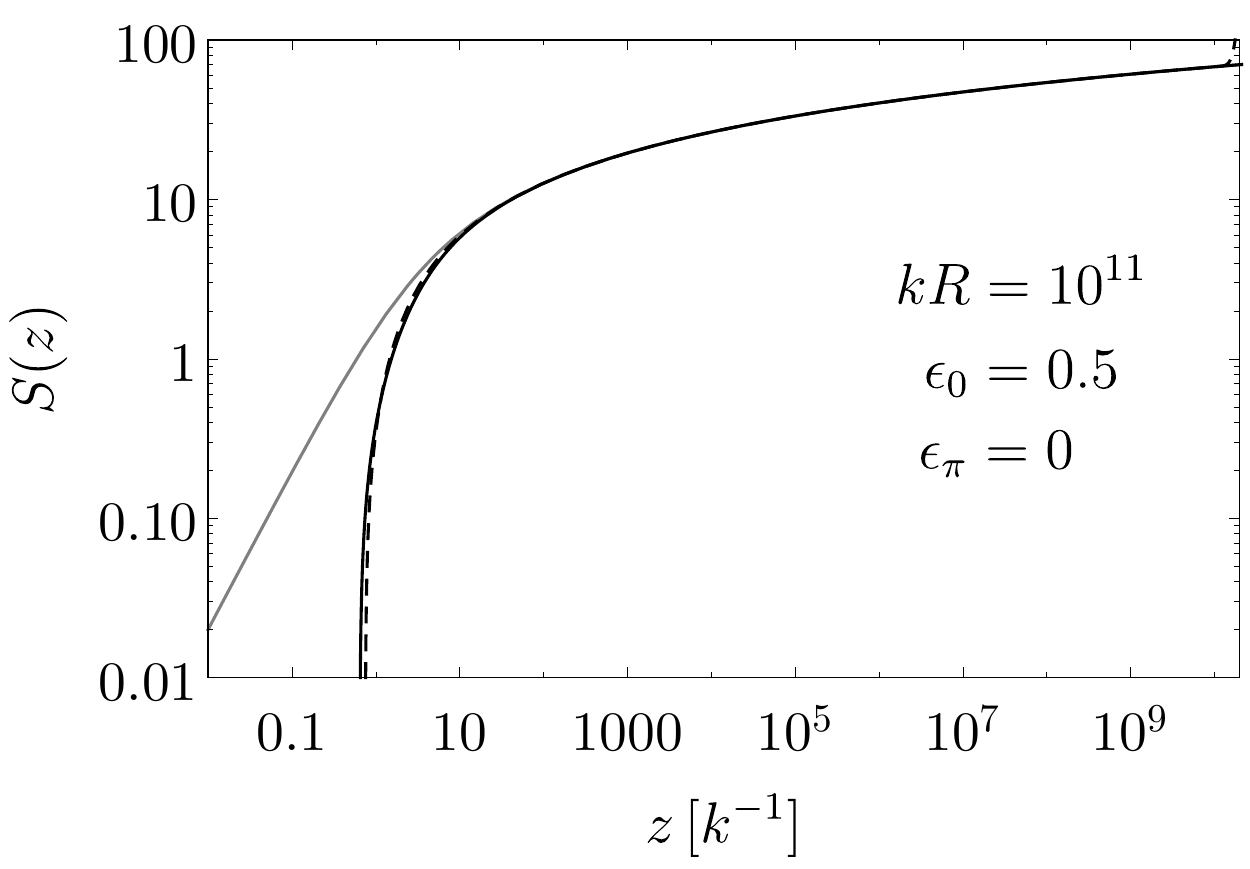}}}}
  \caption{Metric warping (a) and Dilaton profile (b) of the Linear Dilaton model with inflation in the case where the dilaton
  itself serves as a stabilizer. Next to our perturbative results (black) we also show a numerical solution (black, dashed) and the static case solution (gray).
  For the chosen parameters, the resulting Hubble rate is $H\approx8.5\times10^{-18}k\approx8.5\times10^{-15}\,\GeV$ corresponding to a reheating temperature
  around the electroweak scale (assuming maximally efficient reheating).}
  \label{Fig:LD2}
\end{figure}

In the dynamical case, by contrast, the assumption of dilaton stabilization affects the final form of the solution of $f$ and $S$.
The bulk solution is still given perturbatively by the ansatz \eqref{eq:LDAnsatz} with the general solution \eqref{eq:LDSolution}.
As the bulk Lagrangian is fully specified we require that the (55) equation is solved to linear order in $\delta=H^2/k^2$, which is the case 
only if $c_2=0$. As above, the boundary conditions are required to be fulfilled to leading order in $\delta$.
While the boundary conditions for $f$ are unchanged at leading order in $\delta$, 
it is evident that the boundary conditions for $S$ in \eqref{eq:CWJumpConditionLDS} are modified as compared to \eqref{eq:CWJumpConditions}.
Deviations of the brane tensions compared to the static case should again be small, i.e.\ $\eta_0=-1+\epsilon_0$ and $\eta_\pi=1+\epsilon_\pi$
with $\epsilon_{0,\pi}\ll1$.
Finally, we assume that the boundary potentials are strong, $\xi_{0,\pi} M\gg k$,\footnote{%
Since both $\xi$ and $k$ are dilaton shift symmetry breaking parameters, one can control the relative size of $k$ compared to $\xi M$.}
and require again the physical condition $S(0)=0$.
Altogether, a consistent solution is given by
\begin{align}\nonumber
 c_1~&=~\frac{243}{2}\,f_{s,\pi}^3\frac{\epsilon_0+\epsilon_\pi}{\epsilon_0+\epsilon_\pi\,f_{s,\pi}^3}\;,& c_2~&=~0\;, & \\ \label{eq:SolLDS}
 c_3~&=~\frac{27}{2}\left(1-f_{s,\pi}^3\right)\frac{\epsilon_0}{\epsilon_0+\epsilon_\pi\,f_{s,\pi}^3}\;,\quad\text{and}& 
 \delta~&=~\frac43\,\frac{\epsilon_0+\epsilon_\pi\,f_{s,\pi}^3}{f_{s,\pi}^3-1}\;.&
\end{align}
The solution for the metric warping and dilaton profile is fully specified by this. Note that there is no constraint on the relative tensions of the two branes in this case,
meaning that $\epsilon_0$ and $\epsilon_\pi$ can be varied independently.
We note that the last relation of \eqref{eq:SolLDS} can be written as 
\begin{equation}
H^2~=~\frac{\rho_0}{3\,\displaystyle{\left.M_{\mathrm{Pl},\mathrm{eff}}^2\right|^{\mathrm{LD}}_{H=0}}}+\frac{f_{s,\pi}^3\,\rho_\pi}{3\,\displaystyle{\left.M_{\mathrm{Pl},\mathrm{eff}}^2\right|^{\mathrm{LD}}_{H=0}}}\;.
\end{equation}
For the case $\rho_\pi=0$ corresponding to $\epsilon_\pi=0$ the standard expansion law on the visible brane is recovered.
In the following we limit ourselves to this case, i.e.\ we assume that inflation is driven from a surplus energy density located solely on the visible brane.
The solution for the metric warping then is
\begin{equation}
f(z)~=~1+\frac{2\,k}{3}\,|z|+H^2\,\pi^3\,R^3\,k^2\,|z|\,\frac{4\left(-9+6\,k\,|z|+4\,k^2\,z^2\right)}{27\,\left(3+2\,k\,|z|\right)^2}~+~\mathrm{h.o.}\;,
\end{equation}
while the dilaton profile at leading order is given by
\begin{equation}
S(z)~=~3\,\ln\,\left(1+\frac{2\,k}{3}\,|z|\right)- H^2\,\pi^3\,R^3\,k^2\,|z|\,\frac{16\left(3+k\,|z|\right)}{\left(3+2\,k\,|z|\right)^3}~+~\mathrm{h.o.}\;.
\end{equation}
As before $\mathrm{h.o.}$ denotes terms of higher order in $H^2/k^2\ll1$ or lower order in $k\pi R\gg 1$.
The solutions are displayed in \Figref{Fig:LD2} together with a numerical solution. 

Given the metric warping $f(z)$, the effective Planck mass during inflation can be computed to be 
\begin{equation}\label{eq:PlEffLD}
M_{\mathrm{Pl},\mathrm{eff}}^2~=~\left.M_{\mathrm{Pl},\mathrm{eff}}^2\right|^{\mathrm{LD}}_{H=0}\times\left(1+\frac49\,H^2\,k\,\pi^3\,R^3 +\mathrm{h.o.} \right)\;.
\end{equation}
We note that the Planck mass is enhanced during inflation meaning that the strength of gravity is reduced. 
Since inflation here is solely driven from the visible brane, the simplifying assumptions of \cite{Giudice:2002vh} hold and we can
interpret our result in terms of the inflationary tensor-to-scalar ratio. Since gravity is weakened during inflation 
the amplitude of tensor modes is reduced.

There is no upper bound on $H$ from perturbativity requirement on gravity. It is noteworthy, however, that the corrections to the static case are proportional to $(H/k)^2 (k R)^3$ 
in contrast to all other models above, where the $k$ dependence cancels and the corrections were proportional to powers of $(H R)$ only.
This does not modify the conclusion of \cite{Giudice:2002vh}, that the 4D consistency relation $\mathcal{P}_\mathrm{T}(\ell)/\mathcal{P}_\mathrm{S}(\ell)=-4\,n_\mathrm{T}$ 
also holds in the 5D case.

The extra dimension is stabilized by the dilaton at a size $R$, which is determined by
the transcendental relation
\begin{equation}\label{eq:StaticRadiusLD}
 S_\pi-S_0~=~3\,\ln\left(1+\frac23\,k\,\pi\,R\right)-\frac{6\,\epsilon_0}{3+2\,k\,\pi\,R}-2\,\epsilon_\pi\;.
\end{equation}
Clearly, this corresponds to \eqref{eq:StaticRadius} of the static case. For our particular case of interest ($\epsilon_\pi=0$) 
the corrections to the radius with respect to the static case result \eqref{eq:StaticRadius} are $\mathcal{O}(kR)^{-1}$ and, thus, completely negligible.

\subsection{Comparison to results in the literature}
\label{sec:KRcomparion}
Inflation in the Linear Dilaton model has recently also been studied by Kehagias and Riotto (KR) \cite{Kehagias:2016kzt}.
They have investigated inflation in the LD model under the assumption that the Standard Model and the inflaton both reside on the 
UV brane and have found that the tensor modes are suppressed. 
Our setting is conceptually different because we consider the case 
that the visible brane, which hosts the Standard Model and inflaton fields, is the IR brane such that the gauge-hierarchy problem can also be solved.
We also discuss the difference between stabilizing the extra dimension by the dilaton 
compared to the conventional Goldberger-Wise stabilization.
Investigating the setup considered by KR we find qualitative agreement (i.e.\ suppression of the tensor modes)
but we were not able to quantitatively reproduce their results on the metric warping, dilaton profile, and the effective Planck mass.

\section{Conclusions}
\label{sec:Conclusions}

We have considered the scheme of inflation in theories with extra space
dimensions. In this framework some novel questions arise: why do some
dimensions inflate while others are frozen? In the framework of UV-complete
theories (as e.g.\ string theory) this question is related to the mechanism
of moduli stabilization. Another question concerns the location of the
inflaton field (is it a brane- or a bulk-field) and whether the predictions
of the inflationary scenario are influenced by the presence of extra
dimensions. This is the question discussed in the present paper. We have
concentrated our analysis on those situations where the extra dimensions
explain the hierarchy between the weak-scale and the Planck scale. In this
case the relative strength of gravity varies in the bulk between visible and hidden
brane and this can have consequences for the size of inflationary tensor
modes, discussed here in detail. Examples under consideration are large
extra dimensions (LED), the Randall-Sundrum scenario (RS) and the linear
dilaton model (LD).
Up to now the discussion concentrated mainly on simplified cases that
satisfy the IRB assumption where the inflaton sits on the visible 
brane and where the mechanism of stabilizing the extra dimensions is assumed
not to influence the predictions of inflation.

In a first step we have reexamined the IRB case for LED and RS where exact
solutions could be obtained. We observed enhanced tensor modes compared to
inflationary prediction in four space-time dimensions. We also stress that
in these cases we obtain an upper limit on the Hubble scale $H$. The LD case
is more complicated due to the presence of an additional bulk field
(the dilaton). We are not able to find exact solutions here but can derive
a perturbative expansion in $H^2$. Within the LD framework we find that the
naive IRB assumption leads to inconsistencies. Contributions from the hidden
brane (or bulk) are needed to obtain the conventional inflationary scenario.
This leads us to a scheme of ``remote'' inflation, where inflation is 
(partially) driven by energies on the hidden brane.

Motivated by this observation we reconsidered also the RS case beyond the
IRB assumption and the properties of (partially) remote inflation. We
provide a general class of inflationary solutions for remote inflation
that include some specific cases (as NKKK) discussed earlier. Depending
on the specific situations, tensor modes could be enhanced or reduced.
The consequences for the tensor-to-scalar ratio are not known yet as this
would require more calculations beyond the ones given in this paper. In
some cases we find an upper limit on the scale of $H$ similar to that found
in the IRB case. The analysis of remote inflation in the LD case leads
to similar results. The calculation is performed perturbatively in $H^2$ and
supported by a full numerical solution. Still it would be desirable to extend
this to an exact solution as we had derived in the LED and RS case.

The complications in the LD case come from the presence of the additional
dilaton field, and this opens new possibilities. One could use the dilaton
to stabilize the size of the extra dimension within the scheme itself (without
the need for additional stabilizer fields). We have discussed this situation
in detail and found the surprising result that this scheme can be made
consistent with the IRB assumption (with the inflaton field on the IR brane).
In this scheme one obtains a reduced tensor-to-scalar ratio 
while there is no upper limit on $H$.

The presence of extra dimensions can have strong effects on the prediction
of inflationary models, especially in those cases where extra dimensions
provide a solution to the weak-scale hierarchy problem. This is an exciting
situation in view of new observations concerning the tensor-to-scalar
ratio of fluctuations of the cosmic microwave background.

\subsection*{Acknowledgments}
We thank Stefan F\"orste for useful discussions. 
AT also thanks Zhongyi Zhang for useful discussions.
This work has been supported by the German
Science Foundation (DFG) within the SFB-Transregio TR33 ``The Dark Universe".

\appendix
\section{Agreement with earlier results on the RS1 case}
\label{App:RS}
We show that our results in \eqref{eq:HRS}-\eqref{eq:MplEffRS} on the Hubble rate, metric warping, and effective Planck mass for the RS1 case are in
agreement with the results obtained by Giudice et al.\ \cite{Giudice:2002vh}.

In order to make the connection with the results of \cite{Giudice:2002vh} one has to recall that they work in the ``$\pi$-frame'' while we have chosen to put the visible brane at $z=0$ throughout this note. 
In order to reproduce the metric warping in eq.\ (94) of \cite{Giudice:2002vh} one performs the coordinate transformation $z\to-z+\pi R$
in \eqref{eq:WarpFactorRS} to obtain 
\begin{equation}
f^2(z)\left|_{\pi-\text{frame}}\right.~=~\frac{H^2}{\mu^2}\,\mathrm{e}^{\mu\,z}+\left[1+\frac{H^2}{\mu^2}\left(\frac{2\,\omega-1}{\omega}\right)\right]\,\frac{\mathrm{e}^{-\mu\,z}}{\omega}-\frac{2\,H^2}{\mu^2}\;. 
\end{equation}
Using the identities $K\equiv\mu/2$ and $\Omega^2\equiv\omega$ one then finds 
\begin{equation}
 n^2(z)~\equiv~\Omega^2\,f^2(z)|_{\pi-\text{frame}}~=~\frac{\Omega^2\,H^2}{4\,K^2}\,\mathrm{e}^{2\,K\,z}+\left[1+\frac{H^2\left(2\,\Omega^2-1\right)}{4\,\Omega^2\,K^2}\right]\,\mathrm{e}^{-2\,K\,z}-\frac{\Omega^2\,H^2}{2\,K^2}\;, 
\end{equation}
in perfect agreement with \cite{Giudice:2002vh}.
Using the same identities the Hubble rate \eqref{eq:HRS} can be written as
\begin{equation}
 H^2~=~\frac{\epsilon_0\,\mu^2}{2}\,\frac{\omega}{1-\omega}~=~\frac{\kappa^2\,\mu\,\rho_0}{6}\,\frac{\omega}{1-\omega}~\stackrel{\text{\cite{Giudice:2002vh}}}{=}~\frac{K\,\rho_0}{3\,M^3}\,\frac{\Omega^2}{1-\Omega^2}\;.
\end{equation}
This agrees with eq.\ (95) of \cite{Giudice:2002vh} up to a factor of $\Omega^2$ which is simply due to their choice of $M_*\sim M_\mathrm{Pl}$ as 
the fundamental scale. 

\section{Dictionary of different coordinate conventions}
\label{App:Dict}

In the discussion of the Linear Dilaton model multiple coordinate conventions have been used in the literature.
The coordinates in this work are chosen such that the extra dimension is flat implying that $z$ denotes the proper length of the extra dimension. 
A different natural choice of coordinates other than ours is given by
\begin{equation}
 y(z)~=~\frac{3}{2\,k}\,\ln\left(1+\frac23\,k\,z\right)\;,\qquad\Longleftrightarrow\qquad z(y)~=~\frac{3}{2\,k}\left(\e^{2\,k\,y/3}-1\right)\;.
\end{equation}
In this basis the metric of the static case is given by
\begin{equation}
 \mathrm{d}s^2~=~\e^{4\,k\,y/3}\left(\eta_{\mu\nu}\,\mathrm{d}x^\mu\,\mathrm{d}x^\nu-\mathrm{d}y^2\right)\;,
\end{equation}
while the dilaton profile is given by
\begin{equation}
 S(y)~\equiv~\phi(y)~=~2\,k\,y\;.
\end{equation}
The translation of all of our results to this basis is straightforward. In particular, we emphasize
that our solutions \eqref{eq:LDSolution} as well as the specifically determined coefficients \eqref{eq:SolLD}
and \eqref{eq:SolLDS} are stated in a form which is independent of the chosen basis.
Depending on the desired basis the explicit form of the solutions can be obtained by using
$f_{s}(y)=\e^{2\,k\,y/3}$ instead of $f_{s}(z)~=~1+2\,k\,z/3$.

\bibliography{MyLibrary}
\addcontentsline{toc}{section}{Bibliography}
\bibliographystyle{NewArXiv} 
\end{document}